\documentclass[preprint]{elsarticle}

\usepackage{lineno,hyperref}
\usepackage{enumerate}
\usepackage{amssymb}
\usepackage[italicdiff]{physics}
\usepackage{graphicx}
\usepackage{subfigure}
\usepackage{float}
\usepackage{courier}
\usepackage{array, makecell}

\usepackage{listings}
\usepackage{color}
\definecolor{dkgreen}{rgb}{0,0.6,0}
\definecolor{gray}{rgb}{0.5,0.5,0.5}
\definecolor{mauve}{rgb}{0.58,0,0.82}
\modulolinenumbers[5]

\begin{document}

\begin{frontmatter}

\title{An Improved Framework of GPU Computing for CFD Applications on Structured Grids using OpenACC}

\author{Weicheng Xue\corref{cor1}\fnref{fn1}}
\ead{weich97@vt.edu}
\author{Charles W. Jackson\fnref{fn2}}
\ead{cwj5@vt.edu}
\author{Christoper J. Roy\fnref{fn3}}
\ead{cjroy@vt.edu}

\cortext[cor1]{Corresponding author}
\fntext[fn1]{Graduate Research Assistant}
\fntext[fn2]{Graduate Research Assistant}
\fntext[fn3]{Professor}
\address{Virginia Tech Kevin T. Crofton Department of Aerospace and Ocean Engineering,\\
	215 Randolph Hall, Blacksburg, VA 24061, US}

\begin{abstract}
This paper is focused on improving multi-GPU performance of a research CFD code on structured grids. MPI and OpenACC directives are used to scale the code up to 16 GPUs. This paper shows that using 16 P100 GPUs and 16 V100 GPUs can be 30$\times$ and 70$\times$ faster than 16 Xeon CPU E5-2680v4 cores for three different test cases, respectively. A series of performance issues related to the scaling for the multi-block CFD code are addressed by applying various optimizations. Performance optimizations such as the pack/unpack message method, removing temporary arrays as arguments to procedure calls, allocating global memory for limiters and connected boundary data, reordering non-blocking MPI I\_send/I\_recv and Wait calls, reducing unnecessary implicit derived type member data movement between the host and the device and the use of GPUDirect can improve the compute utilization, memory throughput, and asynchronous progression in the multi-block CFD code using modern programming features.
\end{abstract}


\newpageafter{abstract}

\end{frontmatter}


\section{Introduction}

Computational Fluid Dynamics (CFD) is a method to solve problems related to fluids numerically. There are numerous studies applying a variety of CFD solvers to solve different fluid problems. Usually these problems require the CFD results to be generated quickly as well as precisely. However, due to some restrictions of the CPU compute capability, system memory size or bandwidth, highly refined meshes or computationally expensive numerical methods may not be feasible. For example, it may take thousands of CPU hours to converge a 3D Navier-Stokes flow case with more than millions of degrees of freedoms. In such a circumstance, high performance parallel computing~\cite{parallelcomputing} enables us to solve the problem much faster. Also, parallel computing can provide more memory space (either shared or distributed) so that large problems can be solved.

Parallel computing differs from serial computing in many aspects. On the hardware side, a parallel system commonly has multi/many-core processors or even accelerators such as GPUs, which enable programs to run in parallel. Memory in a parallel system is either shared or distributed~\cite{parallelcomputing}, with unified memory address~\cite{landaverde2014investigation} and non-unified memory address usually being used, respectively. On the software side, there are various programming models for parallel computing including OpenMP~\cite{OpenMP2}, MPI~\cite{MPI2}, CUDA~\cite{CUDA}, OpenCL~\cite{OpenCL} and OpenACC~\cite{OpenACC}. Different parallel applications can utilize different parallel paradigms based on a pure parallel model or even a hybrid model such as MPI+OpenMP, MPI+CUDA, MPI+OpenACC, OpenMP+OpenACC, etc.

For multi/many-core computing, OpenMP, MPI and hybrid MPI+OpenMP have been widely used and their performance has also been frequently analyzed in various areas, including CFD. Gourdain et al.~\cite{gourdain2009high,gourdain2009high2} investigated the effect of load balancing, mesh partitioning and communication overhead in their MPI implementation of a CFD code, on both structured and unstructured meshes. They achieved good speedups for various cases up to thousands of cores. Amritkar et al.~\cite{amritkar2014efficient} pointed out that OpenMP can improve data locality on a shared memory platform compared to MPI in a fluid-material application. However, Krpic et al.~\cite{krpic2012green} showed that OpenMP performs worse when running large scale matrix multiplication even on shared-memory computer system when compared to MPI. Similarly, Mininni et al.~\cite{mininni2011hybrid} compared the performance of the pure MPI implementation and the hybrid MPI+OpenMP implementation of an incompressible Navier-Stokes solver, and found that the hybrid approach does not outperform the pure MPI implementation when scaling up to about 20,000 cores, which in their opinion may be caused by cache contention and memory bandwidth. In summary, it can be concluded that MPI is more suitable for massively parallel applications as it can help achieve better performance compared to OpenMP.

In addition to accelerating a code on the CPU, accelerators such as GPU~\cite{owens2008gpu} are becoming popular in the area of scientific computing. CUDA~\cite{CUDA}, OpenCL~\cite{OpenCL}, and OpenACC~\cite{OpenACC} are the three commonly used programming models for the GPU. CUDA and OpenCL are mainly C/C++ extensions (CUDA has also been extended to Fortran) while OpenACC is a compiler directive based interface, therefore CUDA and OpenCL are more troublesome in terms of programming, requiring a lot of user intervention. CUDA is proprietary to NVIDIA and thus can only run on NVIDIA GPUs. OpenCL supports various architectures but it is a very low level API, which is not easy for domain scientists to adapt to. Also, although OpenCL has a good portability across platforms, a code may not run efficiently on various platforms without specific performance optimizations and tuning. OpenACC has some advantages over CUDA and OpenCL. Users only need to add directives in their codes to expose enough parallelisms to the compiler which determines how to accelerate the code. In such a way, a lot of low level implementation can be avoided, which provides a relatively easy way for domain scientists to accelerate their codes on the GPU. Additionally, OpenACC can perform fairly well across different platforms even without significant performance tuning. However, OpenACC may not reveal some parallelisms if there is a lack of performance optimizations. Therefore, OpenACC is usually assumed to be slower than CUDA and OpenCL, but it is still fairly fast. Even for some occasions, OpenACC can be the fastest~\cite{herdman2012accelerating}, which is surprising. To program on multiple GPUs, MPI may be needed, i.e., the MPI+OpenACC hybrid model may be required. CPUs are set as hosts and GPUs are set as accelerator devices, which is referred to as the offload model, in which the most computational expensive portion of the code is offloaded to the GPU, while the CPU handles instructions of controls and file I/O.

A lot of work has been done to leverage GPUs for CFD applications. Jacobsen et al.~\cite{jacobsen2013multi} investigated multi-level parallelisms for the classic incompressible lid driven cavity problem on GPU clusters using MPI+CUDA and hybrid MPI+OpenMP+CUDA implementations. They found that the MPI+CUDA implementation performs much better than the pure CPU implementation but the hybrid performs worse than the MPI+CUDA implementation. Elsen et al.~\cite{elsen2008large} ported a complex CFD code to a single GPU using BrookGPU~\cite{buck2004brook} and achieved a speedup of 40$\times$ for simple geometries and 20$\times$ for complex geometries. Brandvik et al.~\cite{brandvik2008acceleration} applied CUDA to accelerate a 3D Euler problem using a single GPU and got a speedup of 16$\times$. Luo et al.~\cite{luo2013performance} applied MPI+OpenACC to port a 2D incompressible Navier-Stokes solver to 32 NVIDIA C2050 GPUs and achieved a speedup of 4$\times$ over 32 CPUs. They mentioned that OpenACC can increase the re-usability of the code due to OpenACC's similarity to OpenMP. Xia et al.~\cite{xia2015openacc} applied OpenACC to accelerate an unstructured CFD solver based on a Discontinuous Galerkin method. Their work achieved a speedup of up to 24$\times$ on one GPU compared to one CPU core. They also pointed out that using OpenACC requires the minimum code intrusion and algorithm alteration to leverage the computational power of GPU. Chandar et al.~\cite{chandar2013hybrid} developed a hybrid multi-CPU/GPU framework on unstructured overset grids using CUDA. Xue et al.~\cite{xue2018multi} applied multiple GPUs for a complicated CFD code on two different platforms but the speedup was not satisfactory (only up to 4$\times$ on a NVIDIA P100 GPU), even with some performance optimizations. Also, Xue et al.~\cite{xue2020multi} investigated the multi-GPU performance and its performance optimization of a 3D buoyancy-driven cavity solver using MPI and OpenACC directives. They showed that decomposing the total problem in different dimensions affects the strong scaling performance significantly when using multiple GPUs. Xue et al.~\cite{xue2020heterogeneous} further applied the heterogeneous computing to accelerate a complicated CFD code on a CPU/GPU platform using MPI and OpenACC. They achieved some performance improvements for some of their test cases, and pointed out the communication and synchronization overhead between the CPU and GPU may be the performance bottleneck. Both of the works in Ref~\cite{chandar2013hybrid, xue2020heterogeneous} showed that the hybrid CPU/GPU framework can outperform the pure GPU framework to some degree, but the performance gain depends on the platform and application.

\section{Description of the CFD code: SENSEI}

SENSEI (Structured, Euler/Navier-Stokes Explicit-Implicit Solver) is our in-house 2D/3D flow solver initially developed by Derlaga et al~\cite{derlaga2013sensei}, and later extended to a turbulence modeling code base through an object-oriented programming manner by Jackson et al.~\cite{jackson2019turbulence} and Xue et al.~\cite{xue2020code}. SENSEI is written in modern Fortran and is a multi-block finite volume CFD code. An important reason of why SENSEI uses structured grid is that the quality of mesh is better using a multi-block structured grid than using an unstructured grid. In addition, memory can be used more efficiently to obtain better performance since the data are stored in a structured way in memory. The governing equations can be written in weak form as 
\begin{equation}
\label{governing}
\frac{\partial }{\partial t}\int_\Omega \vec{Q}{\rm d}\Omega +\oint_{\partial \Omega} (\vec{F_{i,n}}-\vec{F_{\nu,n}}){\rm d}A= \int_\Omega \vec{S}{\rm d}\Omega
\end{equation}
where $\vec{Q}$ is the vector of conserved variables, $\vec{F_{i,n}}$ and $\vec{F_{\nu,n}}$ are the inviscid and viscous flux normal components (the dot product of the 2nd order flux tensor and the unit face normal vector), respectively, given as
\begin{equation}
\vec{Q}=
\begin{bmatrix}
\rho\\\rho u\\\rho v\\\rho w\\\rho e_t
\end{bmatrix}, \,
\vec{F_{i,n}}=
\begin{bmatrix}
\rho V_n\\\rho u V_n + n_x p\\\rho v V_n + n_y p\\\rho w V_n + n_z p\\\rho h_t V_n
\end{bmatrix}, \,
\vec{F_{\nu,n}}=
\begin{bmatrix}
0\\n_x \tau_{xx} + n_y \tau_{xy} + n_z \tau_{xz}\\n_x \tau_{yx} + n_y \tau_{yy} + n_z \tau_{yz}\\n_x \tau_{zx} + n_y \tau_{zy} + n_z \tau_{zz}\\
n_x \Theta_{x} + n_y \Theta_{y} + n_z \Theta_{z}
\end{bmatrix}	    
\end{equation}
$\vec{S}$ is the source term from either body forces, chemistry source terms, or the method of manufactured solutions~\cite{oberkampf2010verification}. $\rho$ is the density, $u$, $v$, $w$ are the Cartesian velocity components, $e_t$ is the total energy, $h_t$ is the total enthalpy, $V_n = n_x u+ n_y v + n_z w$ and the $n_i$ terms are the components of the outward-facing unit normal vector. $\tau_{ij}$ are the viscous stress components based on Stokes's hypothesis. $\Theta_i$ represents the heat conduction and work from the viscous stresses. In this paper, both the Euler and laminar Navier-Stokes solvers of SENSEI are ported to the GPU, but not for the turbulence models as the turbulence implementation involves a lot of object-oriented programming features such as overloading, polymorphism, type-bound procedures, etc. These newer features of the language are not supported well by the PGI compiler used, as they may require the GPU to jump from an address to a different address in runtime, which should be avoided when programming on GPUs.

In SENSEI, ghost cells are used for multiple purposes. First, boundary conditions can be enforced in a very straightforward way. There are different kinds of boundaries in SENSEI, such as slip wall, non-slip wall, supersonic/subsonic for inflow/outflow, farfield, etc. Second, from the perspective of parallel computing, ghost cells for connected boundaries contain data from the neighboring block used during a syncing routine so that every block can be solved independently. SENSEI uses pointers of a derived type to store the neighboring block information easily. Unless otherwise noted, all of the results presented here will be using a second-order accurate scheme. Second order accuracy is achieved using the MUSCL scheme~\cite{van1979towards}, which calculates the left and right state for the primitive variables on each face of all cells. Time marching can be accomplished using an explicit M-step Runge-Kutta scheme~\cite{jameson1981numerical} and an implicit time stepping scheme~\cite{ascher1997implicit, kennedy2016diagonally, wu1990three}. In this paper, only the explicit M-step Runge-Kutta scheme is used as the implicit scheme uses a completely objected-oriented way of programming which includes overloading of type-bound procedures.

Even though derived types are used frequently in SENSEI, to promote coalesced memory access and improve cache reuse, struct-of-array (SOA) instead of array-of-struct (AOS) is chosen for SENSEI. This means that, for example, the densities in each cell are stored in contiguous memory locations instead of all of the degrees of freedom for a cell being stored together. Using SOA produces a coalesced memory access pattern which performs well on GPUs and is recommended by NVIDIA~\cite{CUDA}.

SENSEI has the ability to approximate the inviscid flux with a number of different inviscid flux functions. Roe's flux difference splitting~\cite{roe1981approximate}, Steger-Warming flux vector splitting~\cite{steger1981flux}, and Van Leer's flux vector splitting~\cite{van1997flux} are available. The viscous flux is calculated using a Green's theorem approach and requires more cells to be added to the inviscid stencil. For more details on the theory and background see Derlaga et at.~\cite{derlaga2013sensei}, Jackson et al.~\cite{jackson2019turbulence} and Xue et al.~\cite{xue2020code}.

\section{Overview of CPU/GPU Heterogeneous System, MPI and OpenACC}

\subsection{CPU/GPU Heterogeneous System}
\begin{figure}[H]
	\centering
	\includegraphics[width=.57\textwidth]{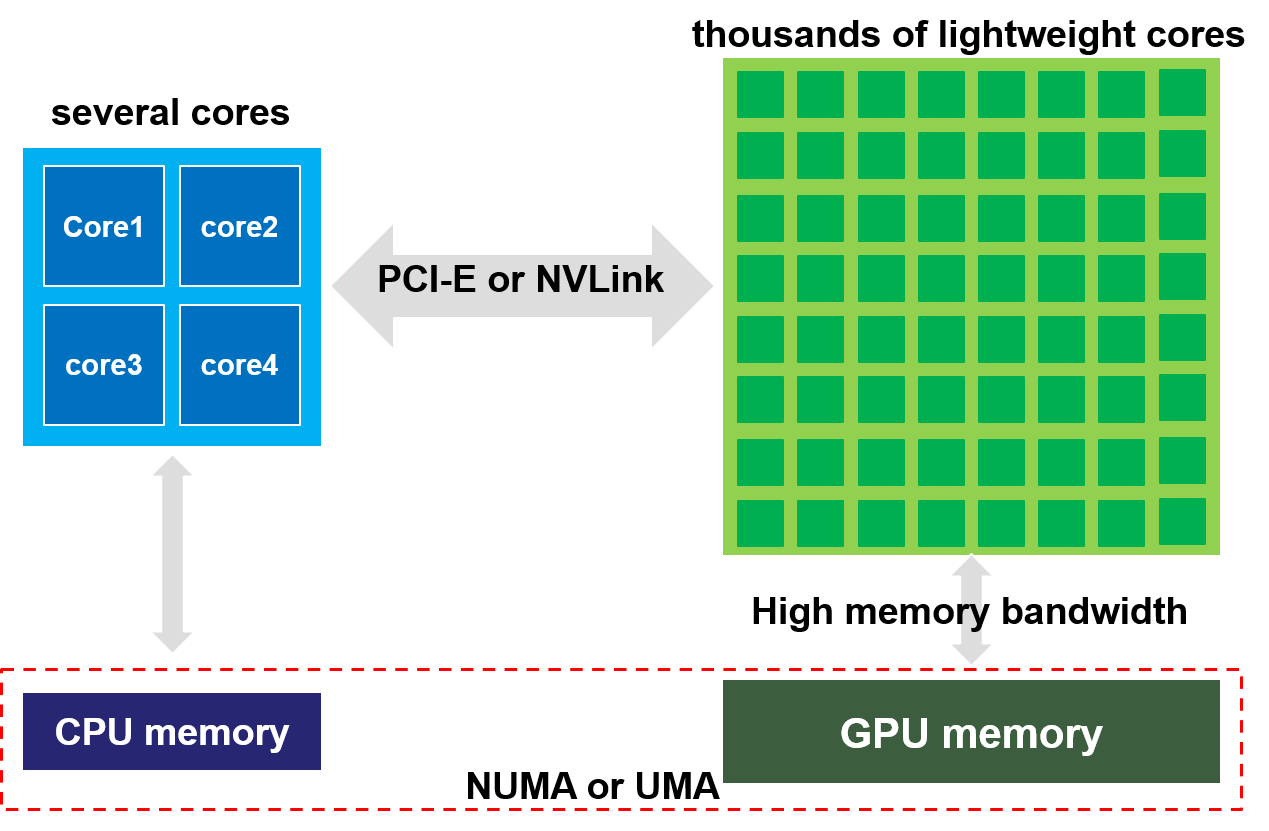}
	\caption{CPU and GPU}
	\label{CPU_GPU}
\end{figure}

\begin{figure}[H]
	\centering
	\includegraphics[width=.57\textwidth]{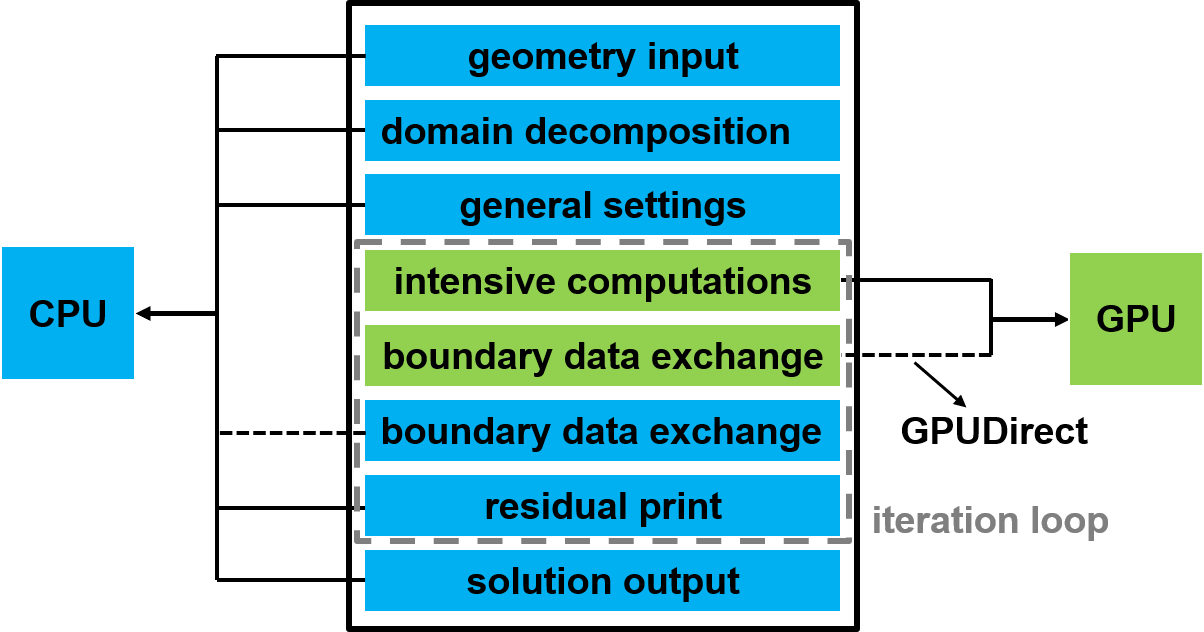}
	\caption{The offload model}
	\label{offload}
\end{figure}

As can be seen in Fig.~\ref{CPU_GPU}, the NVIDIA GPU has more lightweight cores than the CPU, so the compute capability of the GPU is much higher than the CPU. Also, the GPU has higher memory bandwidth and lower latency to its memory. The CPU and the GPU have discrete memories so there are data movements between them, which can be realized through the PCI-E or NVLink. The offload model is commonly used for the pure GPU computing, which can be seen in Fig.~\ref{offload}. In CFD, the CPU deals with the geometry input, domain decomposition and some general settings. Then, the CPU offloads the intensive computations to the GPU. The boundary data exchange can happen either on the CPU or the GPU, depending on whether the GPUDirect is used or not. After the GPU finishes the computation, it moves the solution to the CPU. The CPU finally outputs the solution to files. To obtain good performance, there should be enough GPU threads running concurrently. Using CUDA~\cite{CUDA} or OpenACC~\cite{OpenACC}, there are three levels of tasks: grid, thread block and thread. Thread blocks can be run asynchronously in multiple streaming multiprocessors (SMs) and the communication between thread blocks is expensive. Each thread block has a number of threads. There is only lightweighted synchronization overhead for all threads in a block. All threads in a thread block can run in parallel in the Single Instruction Multiple Threads (SIMT) mode~\cite{SIMT}. A kernel is launched as a grid of thread blocks. Several thread blocks can share a same SMs but all the resources need to be shared. Each thread block contains multiple 32 thread warps. Threads in a warp can be executed concurrently on a multiprocessor. In comparison to the CPU, which is often optimized for instruction controls and for low latency access to cached data, the GPU is optimized for data parallel and high throughput computations.

\subsection{MPI}
MPI (Message Passing Interface) is a programming model for parallel computing~\cite{MPI} which enables data to be exchanged between processors via messages. It can be used on both distributed and shared systems. MPI supports point-to-point communication patterns as well as group communications. MPI also supports the customization of derived data type so transferring data between different processors is easier. It should be noted that a customized derived type may not guarantee fast data transfers. MPI supports the use of C/C++ and Fortran. There are many implementations of MPI including Open MPI~\cite{openmpi} and MVAPICH2~\cite{mvapich2}.

\subsection{OpenACC}
OpenACC is a standard for parallel programming on heterogeneous CPU/GPU systems~\cite{OpenACC}. Very similar to OpenMP~\cite{OpenMP2}, OpenACC is also directive based, so it requires less code intrusion to the original code base compared with CUDA~\cite{CUDA} or OpenCL~\cite{OpenCL}. OpenACC usually does not provide competitive performance compared to CUDA~\cite{hoshino2013cuda, baig2020accelerating, artigues2020evaluation}, however the performance it provides can still satisfy many needs. Compilers such as PGI~\cite{pgi} and GCC can support OpenACC in a way that the compiler detects the directives in a program and decides how to parallelize loops by default. The compiler also handles moving the data between discrete memory locations, but it is the users' duty to inform the compiler to do so. Users can provide more information through the OpenACC directives to attempt to optimize performance. These optimizations will be the focus of this paper.

\section{Domain Decomposition}
 
There are many strategies to decompose a domain, such as using Cartesian~\cite{hager2010introduction, farber2016parallel} or graph topology~\cite{hendrickson2000graph}. Because SENSEI is a structured multi-block code, Cartesian block splitting will be used. With Cartesian block splitting, there is a tradeoff between decomposing the domain in more dimensions (e.g. 3D or 2D domain decomposition) and fewer dimensions (e.g. 1D domain decomposition). The surface area-volume ratio is larger if decomposing the domain in fewer dimensions, which means more data needs to be transferred between different processors. Also, decomposing the domain in 1D can generate slices that are too thin to support the entire stencil when decomposing the domain into many sub-blocks. However, the fewer number of dimensions being composed means that each block needs to communicate with a fewer number of neighbors, reducing the number of transfers and their corresponding latency. 

\begin{figure}[H]
	\centering
	\includegraphics[width=.8\textwidth]{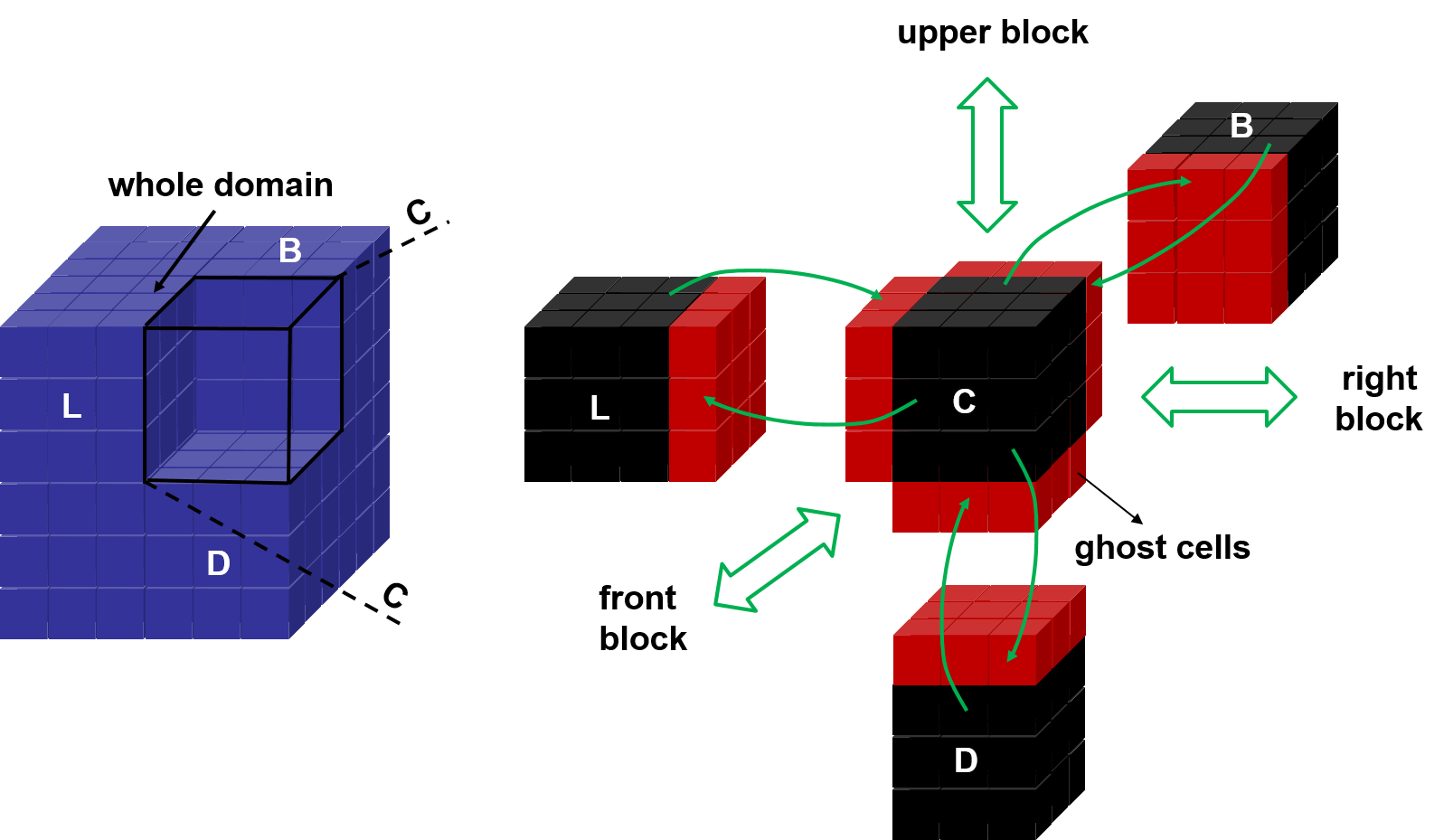}
	\caption{A 3D domain decomposition}
	\label{3DDecomposition}
\end{figure}

By default, SENSEI uses a general 3D or 2D domain decomposition (depending on whether the problem is 3D or 2D) but can switch to 1D domain decomposition if specified. An example of the 3D decomposition is shown in Fig.~\ref{3DDecomposition}. The whole domain is decomposed into a number of blocks. Each block connects to 6 neighboring blocks, one on each face. For each sub-iteration step (as the RK multi-step scheme is used), neighboring decomposed blocks need to exchange data with each other, in order to fill their own connected boundaries. Since data layout of multi-dimensional arrays in Fortran is column-majored, we always decompose the domain starting form the most non-contiguous memory dimension. For example, since the unit stride direction of a three-dimensional array $A(i, j, k)$ is the first index ($i$), $i$ is the last decomposed dimension and $k$ is the first decomposed dimension.

The 3D domain decomposition method shown in Fig.~\ref{3DDecomposition} is a processor clustered method. This method is designed for the scenarios in which the number of processors ($np$) is greater than the number of parent blocks ($npb$), i.e., the number of blocks before the domain decomposition. There are several advantages with this decomposition strategy. First, this method is an "on the fly" approach, which is convenient to use and requires no manual operation or preprocessing of the domain decomposition. Second, it is very robust that it can handle most situations if $np$ is greater than or equal to $npb$. Third, the communication overhead is small due to the simple connectivity, making the MPI communication implementation easy. The load can be balanced well if $np$ is significantly larger than $npb$. Finally, some domain decomposition work can be done in parallel, although the degree may vary for various scenarios.

\begin{figure}[H]
	\centering
	\includegraphics[width=.4\textwidth]{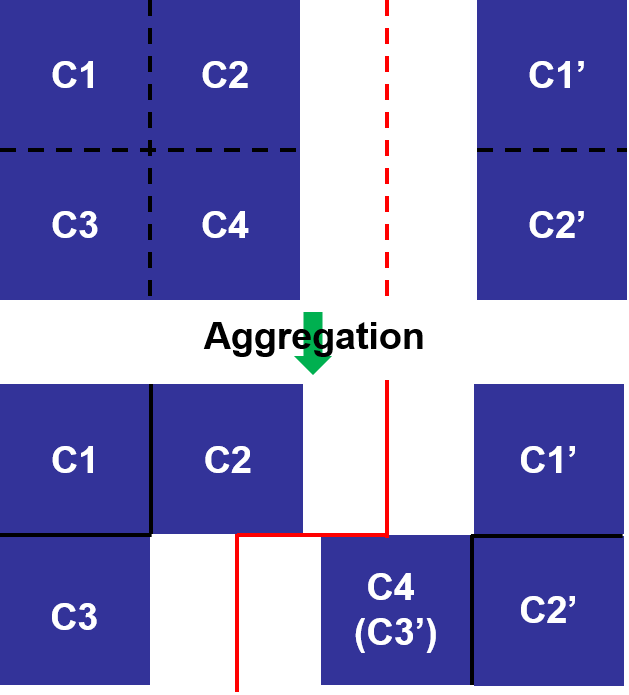}
	\caption{An example showing the domain aggregation}
	\label{Aggregation}
\end{figure}

This domain decomposition method may have load imbalance issue if $np$ is not obviously greater than $npb$, which can be addressed using a domain aggregation technique, similar to building blocks. A simple 2D example of how the domain aggregation works is given in Fig.~\ref{Aggregation}. In this example, the first parent block has twice as many cells as the second parent block. If only two processors are used, the workload cannot be balanced well without over-decomposition and aggregation. With over-decomposition, the first parent block is decomposed into 4 blocks, and one of these decomposed blocks is assigned to the second processor so both processor have the same amount of work to do. It should be noted that the processor boundary length becomes longer due to the processor boundary deflection increasing the amount of communication required. Using the domain aggregation approach, any decomposed block is required to exchange data with its neighbours on the same processor but this does not require MPI communications. Only the new connected boundaries (e.g. the red solid lines in Fig.~\ref{Aggregation}) between neighboring processors need to be updated using MPI routines at each sub-iteration step.

\section{Boundary Decomposition in Parallel and Boundary Reordering}

\begin{figure}[H]
	\centering
	\includegraphics[width=.7\textwidth]{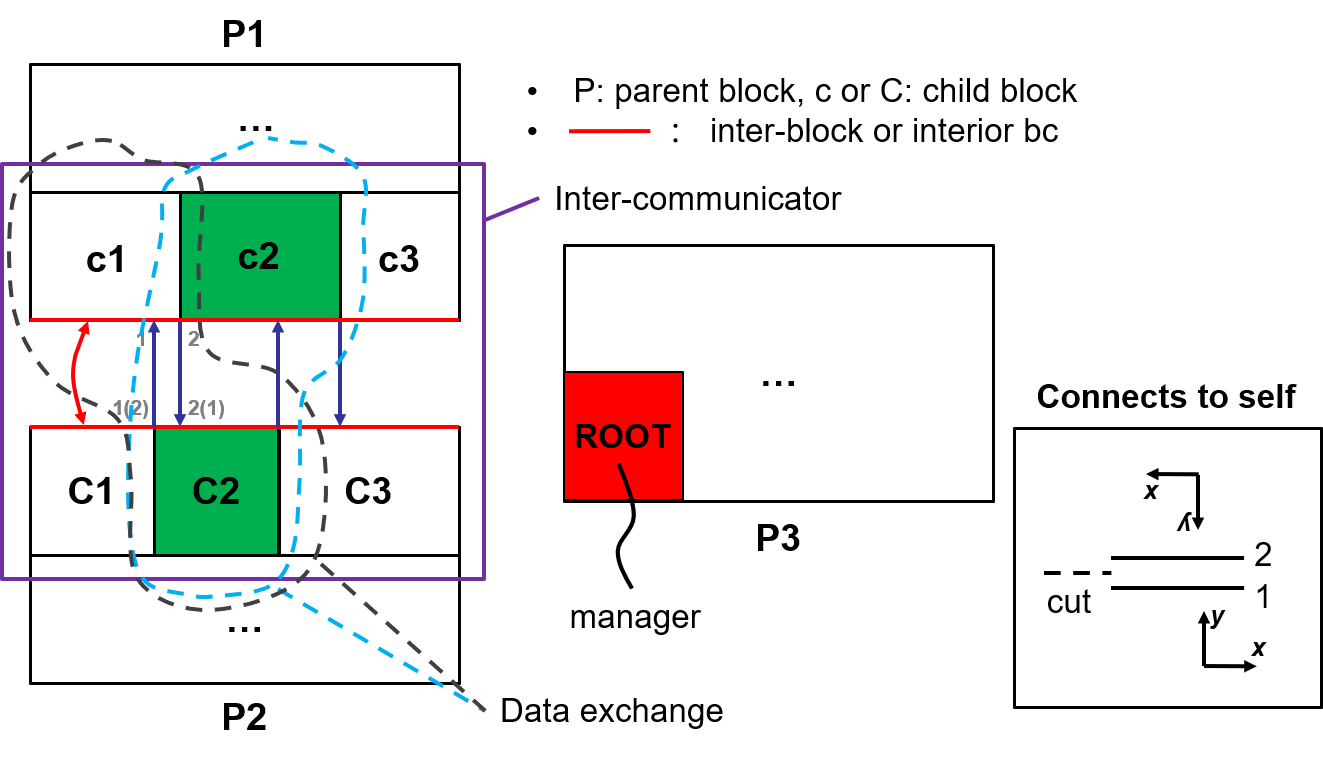}
	\caption{An example of using MPI inter-communicator}
	\label{Inter_Communicator}
\end{figure}

Boundaries also need to be decomposed and updated on individual processors. Initially, only the root processor has all the boundary information for all parent blocks, since root reads in the grid and boundaries. After domain decomposition, each parent block is decomposed into a number of child blocks. These child blocks need to update all the boundaries for themselves. For non-connected boundaries this update is very straightforward as each processor just needs to compare their individual block index range with the boundary index range. For interior boundaries caused by domain decomposition, a family of Cartesian MPI topology routines are used to setup communicators and make communication much less troublesome. However, for connected parent block boundaries, the update (decomposing and re-linking these boundaries) is more difficult, as the update is completed in parallel on individual processors in SENSEI, instead of on the root processor. The parallel process can be beneficial if numerous connected boundaries exist. For every parent block connected boundary, the root processor first broadcasts the boundary to all processors within that parent block and its neighbour parent block, and then returns to deal with the next parent block connected boundary. The processors within that parent block or its neighbour parent block compare the boundary received to their block index ranges. If a processor does not contain any index range of the parent boundary, it moves forward to compare the next parent boundary. Processors in the parent block having this boundary or processors in the neighbour parent block matching part of the neighbour index range are colored but differently. These colored processors will need to update their index range for the connected boundary. To illustrate how we use MPI topology routines and inter-communicators to setup connectivity between neighbour blocks, a 2D example having 3 parent blocks and more than 3 CPUs is given in Fig.~\ref{Inter_Communicator}. Processors which match a parent connected boundary are included in an inter-communicator. The processor in a parent block first sends its index ranges to processors residing in the neighbour communicator. Then a processor in the neighbour communicator matching part of the index range is a neighbour while others in the neighbour communicator not matching the index range are not neighbour processors. Through looping over all the neighbour processors in the neighbour communicator, one processor sets up connectivity with all its connected neighbours. This process is performed in parallel as the root processor does not need to participate in this process except for broadcasting the parent boundary to all processors in the parent block and its neighbour parent block at the beginning. There may be special cases. The first special case is that the root is located at a parent block or its neighbour parent block. The root needs to participate in the boundary decomposition and re-linking process, as shown in Fig.~\ref{Inter_Communicator}. The second special case is given in the lower right square in Fig.\ref{Inter_Communicator}, in which a parent block partly connects to itself, which may make a decomposed block partially connect to itself.

\begin{figure}[H]
	\centering
	\includegraphics[width=.8\textwidth]{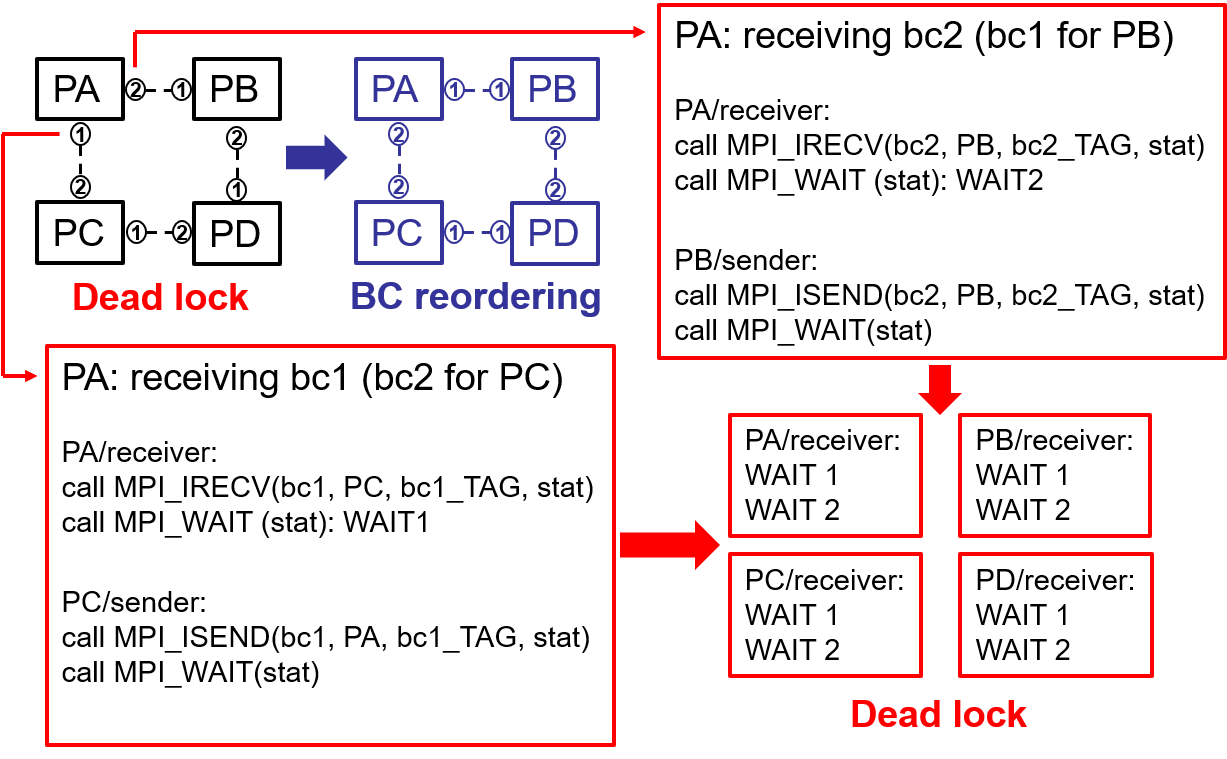}
	\caption{An example of deadlock due to circular dependency}
	\label{deadlock}
\end{figure}

In SENSEI, nonblocking MPI calls instead of blocking calls are used to improve the performance. However, nonblocking MPI calls requires a blocking call such as MPI\_WAIT to finish the communication, and it may cause a deadlock issue for some multi-block cases. An example of the deadlock issue is shown in Fig.~\ref{deadlock}. In this example, there are four processors ($PA$$\sim$$PD$), each with two connected boundaries ($bc1$ and $bc2$). For every processor, it needs to block a MPI\_WAIT call for its $bc1$ to finish first and then for its $bc2$. However, the initial order of boundaries creates a circular dependency issue for all of the processors, and thus no communication can be completed (deadlock). This deadlock issue may happen for both the parent block connections and the child block connections after decomposition. Fig.~\ref{deadlock} shows a solution to the deadlock issue, i.e. reordering boundaries. Therefore, boundary reordering is implemented in SENSEI to automatically deal with such deadlock issues.

\section{Platforms and Metrics}

\subsection{Platforms}
\paragraph{Thermisto}
Thermisto is a workstation in our research lab. It has two NVIDIA Tesla C2075 GPUs and 32 CPU cores. The peak double precision performance is 515 GFLOPS. The compilers used on Thermisto are PGI 16.5 and Open MPI 1.10.0. An compiler optimization of -O4 is used. The GPUs on Thermisto are used mainly for testing and comparison with current generation GPUs.

\paragraph{NewRiver}
NewRiver~\cite{Newriver} is a cluster at Virginia Tech. Each GPU node on NewRiver is equipped with two Intel Xeon E5-2680v4 (Broadwell) 2.4GHz CPUs, 512 GB memory, and two NVIDIA P100 GPUs. Each NVIDIA P100 GPU is capable of up to 4.7 TeraFLOPS of double-precision performance. The NVIDIA P100 GPU offers much higher GFLOPS compared to the NVIDIA C2075 GPU on Thermisto. The compilers used on NewRiver are PGI 17.5 and Open MPI 2.0.0 or MVAPICH2-GDR 2.3b. MVAPICH2-GDR 2.3b is a CUDA-aware MPI wrapper compiler which supports GPUDirect, also available on NewRiver. An compiler optimization of -O4 is used.

\paragraph{Cascades}
Cascades~\cite{Cascades} is another cluster at Virginia Tech. Each GPU node on Cascades is equipped with two Intel Skylake Xeon Gold 3 GHz CPUs, 768 GB memory, and two NVIDIA V100 GPUs. Each NVIDIA V100 GPU is capable of up to 7.8 TeraFLOPS of double-precision performance. The NVIDIA V100 GPU offers the highest GFLOPS among the GPUs we used. The compilers used on Cascades are PGI 18.1 and Open MPI 3.0.0. An compiler optimization of -O4 is used.

\subsection{Performance Metrics}
To evaluate the performance of the parallel code, weak scaling and strong scaling are used. Strong scaling measures how the execution time varies when the number of processors changes for a fixed total problem size, while weak scalability measures how the execution time varies with the number of processors when the problem size on each processor is fixed. Commonly, these two scalings are valuable to be investigated together, as we care more about the weak scaling when we have enough compute resources available to run large problems, while more about the strong scaling when we only need to run small problems. In this paper, since our focus is on the acceleration of the computation and data movement in the iterative solver portion, when measuring productive performance, the timing contribution from the I/O portion (reading in grid, writing out solution) and the one-time domain decomposition is not taken into account.

Two basic metrics used in this paper are parallel speedup and efficiency. Speedup denotes how much faster the parallel version is compared with the serial version of the code, while efficiency represents how efficiently the processors are used. They are defined as follows, 
\begin{equation}
\label{speedup}
\mathrm{speedup}=\frac{t_{serial}}{t_{parallel}}
\end{equation}
\begin{equation}
\label{efficiency}
\mathrm{efficiency}=\frac{\textrm{speedup}}{np}
\end{equation}
where $np$ is the number of processors (CPUs or GPUs).

In order for the performance of the code to be compared well on different platforms and for different problem sizes, the wall clock time per iteration step is converted to a metric called ssspnt (scaled size steps per $np$ time) which is defined in Eq.\ref{ssspnt}.
\begin{equation}
\label{ssspnt}
\mathrm{ssspnt}=s\frac{size \times steps}{np \times time}
\end{equation}
where $s$ is a scaling factor which scales the smallest platform ssspnt to the range of [0,1]. In this paper, $s$ is set to be $10^{-6}$ for all test cases. $size$ is the problem size, $steps$ is the total iteration steps and $time$ is the program solver wall clock time for $steps$ iterations.

\begin{figure}[H]
	\centering 
	\subfigure[ssspnt for super-linear/linear/sub-linear scaling]{ 
		\label{ssspnt1}
		\includegraphics[width=.45\textwidth,trim=5 5 5 5,clip]{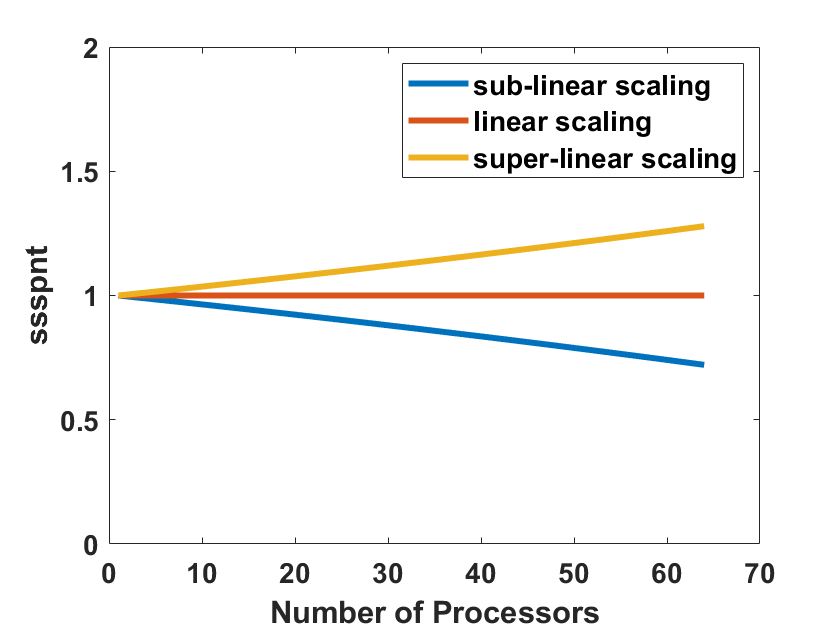} 
	}
	\subfigure[ssspnt for different cases]{ 
		\label{ssspnt2}
		\includegraphics[width=.45\textwidth,trim=5 5 5 5,clip]{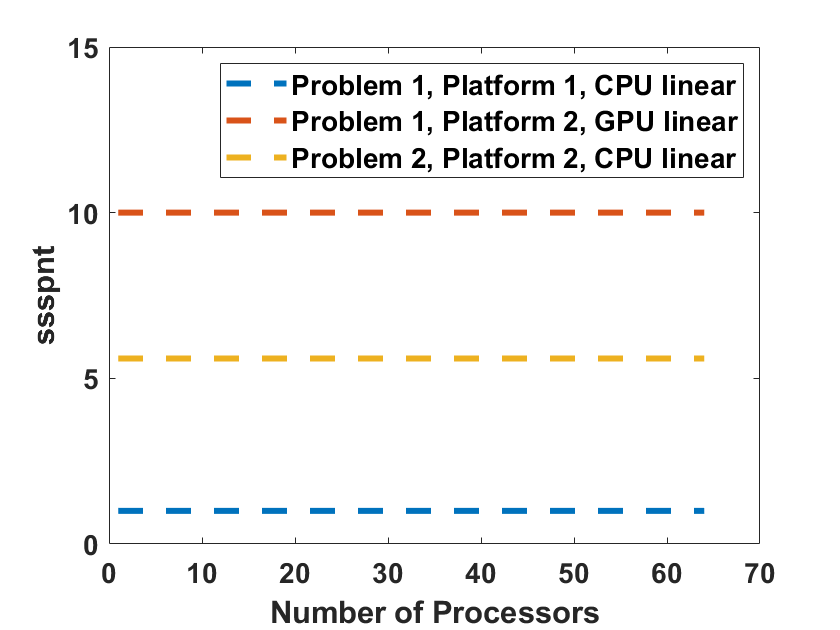} 
	} 
	\caption{An explanation of ssspnt} 
	\label{ssspnt_exp}
\end{figure}

Using ssspnt has some advantages. First, GFLOPS requires knowing the number of operations while ssspnt does not. In most codes, especially complicated codes, it is usually difficult to know the total number of operations. The metric ssspnt is a better way of measuring the performance of a problem than the variable $time$ as $time$ may change if conditions (such as the number of iterations, problem size, etc.) change. Second, using ssspnt is clearer in terms of knowing the relative speed difference under different situations than the metric "efficiency". It is easy to know whether the performance is super-linear or linear or sub-linear, which is shown in Fig.~\ref{ssspnt1}, as well as know the relative performance comparison between different scenarios, which is shown in Fig.~\ref{ssspnt2}. Using ssspnt, different problems, platforms and different scalings can be compared more easily.

Similar to Ref~\cite{xue2020multi}, every $time$ in this paper is measured consecutively for at least three instances. The difference for each $time$ point is smaller than 1\% (usually less than 1 s out of more than 120 s). We also selected a handful of cases to run again to verify the timings were consistent day to day.

\section{OpenACC Parallelization and Optimization}

There is some general guidance for improving the performance of a program on a GPU. First, sufficient parallelism should be exposed to saturate the GPU with enough computational work, that is, the speedup for the parallel portion should compensate for the overhead of data transfers and the parallel setup. Second, the memory bandwidth between the host and the device should be improved to reduce the communication cost, which is affected by the message size and frequency (if using MPI), memory access patterns, etc. It should be noted that all performance optimizations should guarantee the correctness of the implementation. Therefore, this paper proposes and adopts various modifications to increase the speed of various CFD kernels and reduce the communication overhead while always ensuring the correct result is obtained, i.e., the results do not deviate from the serial implementation.

Load balancing, communication overhead, latency, synchronization overhead and data locality are important factors which may affect the performance. The domain decomposition and aggregation methods used in this paper can help solve the load imbalancing issue well; however, the number of dimensions that need to be decomposed may require tuning, especially when given a large number of processors. To reduce the communication overhead of data transfers between the CPU and the GPU, the data should be kept on the GPU as long as possible without being frequently moved to the CPU. Also, non-contiguous data transfer between the CPU and the GPU (large stride memory access) should be avoided to improve the memory bandwidth. To hide latency, kernel execution and data transfer should be overlapped as much as possible, which may require reordering of some portions in the program. To reduce the synchronization overhead, the number of tasks running asynchronously should be maximized. To improve data locality and increase the use of coalesced fetches, data should be loaded into cache as chunks before needed, which can make read and write more efficiently. This paper addresses some of these issues based on profiling outputs.

We should keep in mind that there are some inherent bottlenecks limiting the actual performance of a CFD code on GPUs. Some CFD codes require data exchange to communicate between partitions, which incurs some communication and synchronization overhead. Data fetching in discrete memory may cost more clock cycles than expected due to low actual memory throughput, system latency, etc. Therefore, the actual compute utilization is difficult to increase sometimes and is application dependent. Another limiter of the performance is the need for branching statements in the code. For instance, certain flux functions might execute different branches depending on the local Mach number. This causes threads in a warp to diverge reducing the peak performance possible. The actual speedup after enough performance optimization should still be smaller than the theoretical compute power the GPU can provide. The relation of the actual and theoretical speedup the GPU can provide is not covered in this paper.

\subsection{V0: Baseline}

The baseline GPU version of SENSEI was implemented by McCall~\cite{mccall2017multi},~\cite{mccall2017multilevel}. McCall pointed out that there are some restrictions of the PGI compiler. These restrictions mean the following features cannot be used.
\begin{enumerate}
    \item Procedure optional arguments
    \item Array-valued functions
    \item Multi-dimensional array assignments
    \item Temporary arrays as parameters to a procedure call
    \item Reduction operations on derived type members
    \item Procedure pointers within OpenACC kernels
\end{enumerate}

As can be seen in ~\cite{mccall2017multi} and ~\cite{mccall2017multilevel}, the 1st and 3rd restrictions do not have adverse effects on the performance. The 2nd restriction can be easily resolved by using Fortran subroutines instead of functions. The 5th restriction can be resolved by using scalar variables or arrays instead of derived type members, which has negligible effect on the performance. The 6th restriction can be easily overcome by using the \texttt{select case} or \texttt{if} statements. The 4th restriction indicates that either the compiler needs to automatically generate the temporary arrays or the user should manually create them. However, the temporary arrays deteriorate the code performance significantly. More details about these restrictions can be found in Ref~\cite{mccall2017multi} and \cite{mccall2017multilevel}.

Although the work in Ref~\cite{mccall2017multi} and \cite{mccall2017multilevel} overcame many restrictions to port the code to the GPU, the GPU performance was not satisfactory. A NVIDIA P100 GPU was only 1.3x$\sim$3.4x faster than a single Intel Xeon E5-2680v4 CPU core. which indicates that the GPU was not utilized efficiently. Some performance bottlenecks were fixed in Ref~\cite{xue2018multi}. Profiling-driven optimizations were applied to overcome some performance bottlenecks. First, loops with small sizes were not parallelized as the launch overhead is more expensive than the benefits. As the warp size for NVIDIA GPUs is 32, the compiler may select a thread length of 128 or 256 to parallelize small loops but the loop iteration number for these small loops is less than 10. Second, the kernel of extrapolation to ghost cells was moved from the CPU to the GPU in order to improve the performance, by passing the whole array with indices as arguments. Finally, the kernel of updating corners and edges was parallelized. The eventual speedup of using a single GPU compared to a single CPU was raised to 4.1x for a 3D case on a NVIDIA P100 GPU, but no multi-GPU performance results were shown as the parallel efficiency was not satisfactory.

It should be mentioned that the relative solution differences between the CPU and the GPU code in Ref~\cite{mccall2017multi, mccall2017multilevel, xue2018multi} are much larger than the round-off error mainly due to an incorrect implementation of connected boundary condition and its relevant parallelization. The solution bugs have been fixed in this paper so that the OpenACC framework is extended correctly to multi-block cases. In fact, solution debugging is troublesome using OpenACC, as intermediate results are difficult to check directly on the device. If the data on the GPU side needs to be printed outside of the parallel region, then update of the data on the host side should be made before printing. If the data needs to be known in the parallel region (when a kernel is running), a probe routine which is \texttt{!\$acc routine} type should be inserted into the parallel region to print out the desired data. Keep in mind that both the GPU and CPU have a copy of the data with the same name but in discrete memories.

\begin{figure}[H]
	\centering
	\includegraphics[width=.8\textwidth]{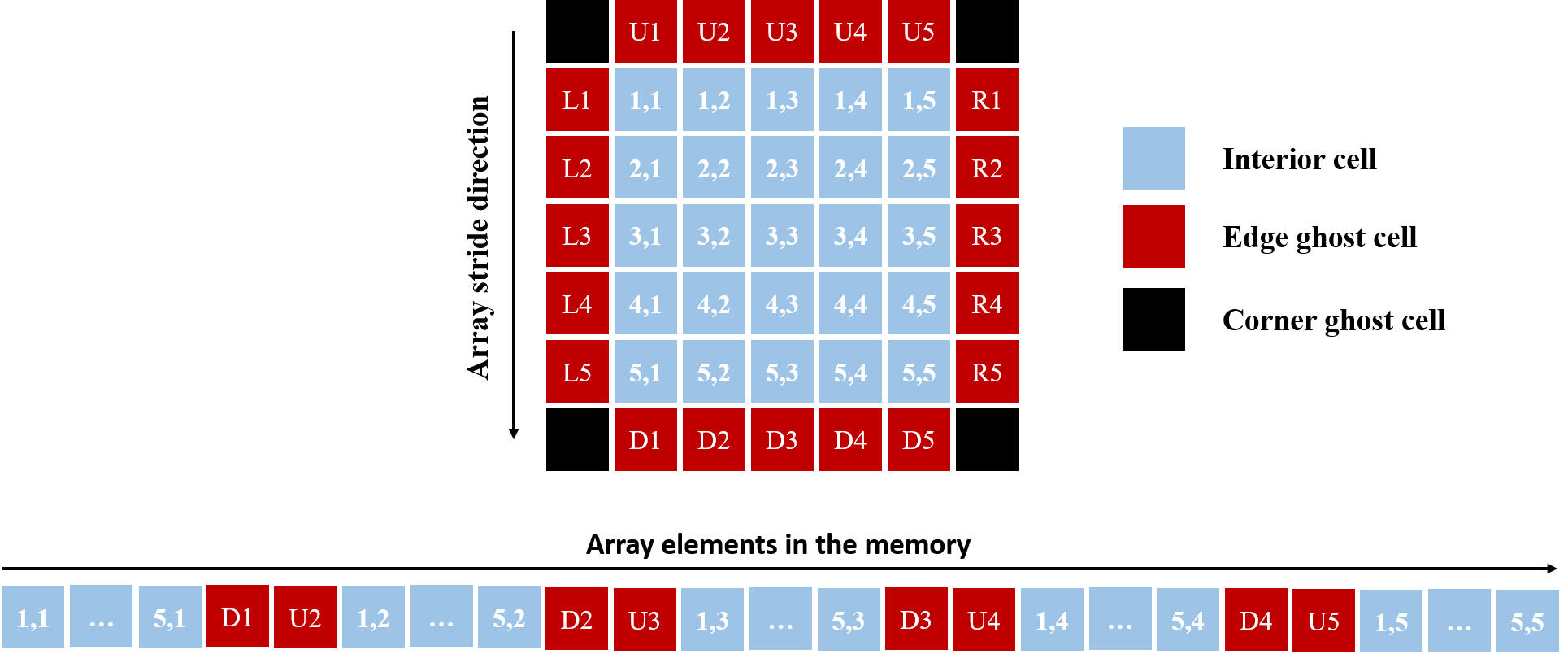}
	\caption{An example of showing ghost cells breaking the non-contiguity of the interior cell}
	\label{conti_nonconti}
\end{figure}

In addition, it should be mentioned that there is a caveat when updating the boundary data between the host and the device using the \texttt{!\$acc update} directive since the ghost cells and interior cells in SENSEI are stored and addressed together, which means that the boundary data are non-contiguous in the memory. Much higher memory throughput can be obtained if the whole piece of data (including the interior cells and boundary cells) instead of array slicing is included. A 2D example can be seen in Fig.~\ref{conti_nonconti}. As Fortran uses column-major storage, the memory stores the array elements by column first. However, the interior cell columns split the ghost cells in memory. For 3D or multi-dimensional arrays, the data layout is more complicated but the principle is similar. If using the method in Listing~\ref{V00_update} to update the boundary data on the device, OpenACC updates the data slice by slice and there are many more invocations. The memory throughput can be about 1/100 to 1/8 of using the method in Listing~\ref{V0_update}, based on the profiling outputs from the NVIDIA visual profiler. In fact, the only implementation difference between the two methods is whether slicing is used or not (in Fortran, array slicing is commonly used), but the performance difference is huge. However, some applications or schemes may require avoiding updating the ghost cell values (due to concerns for solution correctness) at some temporal points when iterating the solver, then a manual data rearrangement, i.e., the pack/unpack optimization, should be applied to overcome the performance deterioration issue.

\begin{lstlisting}[caption={Using slicing to update},label=V00_update, language=Fortran]
! IMIN face update
start_indx = 1 - n_ghost_cells(1)
end_indx   = 2
!$acc update device(
!$acc soln%sblock(blck)%rho(start_indx:end_indx,1:jmax,1:kmax))
!$acc async(1)
\end{lstlisting}

\begin{lstlisting}[caption={Update including ghost cells},label=V0_update, language=Fortran]
! IMIN face update
start_indx = 1 - n_ghost_cells(1)
end_indx   = 2
!$acc update device(
!$acc soln%sblock(blck)%rho(start_indx:end_indx,:,:))
!$acc async(1)
\end{lstlisting}

Table~\ref{V00&V0Profiling} shows the performance of some metrics for the $V00$ (using Listing~\ref{V00_update}) and $V0$ (using Listing~\ref{V0_update}) comparison on the Cascades platform. Using slicing for the \texttt{!\$acc update} directive reduces the memory throughput greatly to about 1\% for the device to host bandwidth and about 8\% for the host to device bandwidth, compared to not using slicing (with ghost cell data included). Also, the total invocations of using slicing is more than 10 times higher than not using slicing. The last row in Table~\ref{V00&V0Profiling} is a reference (NVIDIA profiler reports different fractions of low memory throughput data transfers for different code versions) to show the more serious low memory throughput issue in $V00$. We will show some performance optimizations based on the $V0$ version next, even though $V0$ has larger solution errors than the round-off errors for some cases.
\begin{table}[H]
	\caption{Comparison of V00 and V0 performance metrics}
    \centering
    \label{V00&V0Profiling}
	\begin{tabular}{ccc}
		\hline
		Metrics                   & V00         & V0            \\ \hline
		Device to Host bandwidth, GB/s    & 0.132       & 10.43         \\
		Host to Device bandwidth, GB/s    & 0.9         & 7.21          \\
		Total invocations, times  & 262144      & 20876         \\
		Compute Utilization, \%   & 4.2         & 29.6          \\
        Low memory throughput     & \makecell{124 MB/s for 96.4\%   \\
                                    data transfers}
                                  & \makecell{83.88 MB/s for 11.1\% \\
                                    data transfers}             \\ \hline

	\end{tabular}
\end{table}

\subsection{GPU Optimization using OpenACC}

Although parallelized using the GPU in Ref~\cite{mccall2017multi,mccall2017multilevel} and optimized in Ref~\cite{xue2018multi}, the speedup for SENSEI is still not satisfactory due to some performance issues. The NVIDIA Visual Profiler is used to detect various performance bottlenecks. The bottlenecks include low memory throughput, low GPU occupancy, inefficient data transfers, etc. Different architectures and problems may show different behaviours, which is one of our interests. Second, previously the boundary data needed to be transferred to the CPU first in order to exchange data. We will apply GPUDirect to enable data transfers directly between GPUs.

\paragraph{V1: Pack/Unpack}
The goal of this optimization is to improve the memory throughput and reduce the communication cost if the required data are not located sequentially in memory~\cite{xue2020multi}. As Fortran is a column-majored language, the first index $i$ of a matrix $A(i,j,k)$ denotes the fastest change. A decomposition in the $i$ index direction can generate chunks of data (at $j-k$ planes) which are highly non-contiguous. Decomposing in the $j$ index direction can also cause non-contiguous data transfers. Therefore, the optimization is targeted at solving this issue by converting the non-contiguous data into a temporary contiguous array in parallel using \texttt{loop for} directives and then updating this temporary array between hosts and devices using \texttt{update} directives. Performance gains will be obtained as the threads in a warp can access a contiguous aligned memory region, that is, coalesced memory access is deployed instead of strided memory access. The procedure can be summarized as follows:
\begin{enumerate}
\item Allocate send/recv buffers for boundary cells on $j-k$ planes on devices and hosts if decomposition happens in the $i$ dimension, as the non-contiguous data on $i$ planes make data transfer very slow.
\item Pack the noncontiguous block boundary data to the send buffer, which can be explicitly parallelized using \texttt{!\$acc loop} directives, then update the send buffer on hosts using \texttt{!\$acc update} directives.
\item Have hosts transfer the data through nonblocking MPI\_Isend/MPI\_Irecv calls and blocking MPI\_Wait calls.
\item Update the recv buffer on devices using OpenACC update device directives and finally unpack the contiguous data stored in recv buffer back to noncontiguous memory on devices, which can also be parallelized.
\end{enumerate}

We will show that although extra memory is required for buffers, the memory throughput can be improved to a level similar to that in $V0$ (but $V0$ has larger simulation errors due to the incorrect use of \texttt{!\$acc update}, especially for cases having connected boundaries). Using $V1$, only the boundary data on the $i$ boundary faces are packed/unpacked as such data are highly noncontiguous. The boundary data on the $j$ and $k$ plane are not buffered. The pack/unpack can be parallelized using \texttt{!\$acc loop} directives so that the computational overhead is very small, which can be seen in Listing~\ref{pack_unpack}.

\begin{lstlisting}[caption={A pseudo code of showing how to pack/unpack},label=pack_unpack, language=Fortran]
! IMIN face update
start_indx = 1 - n_ghost_cells(1)
end_indx   = interior_cells
!$acc parallel present(soln, soln%sblock,      &
!$acc                  soln%sblock(blck)%rho,  &
!$acc                  soln%sblock(blck)%vel,  &
!$acc                  soln%sblock(blck)%p,    &
!$acc                  soln%sblock(blck)%temp, &
!$acc                  rho_buffer, vel_buffer, &
!$acc                  p_buffer, temp_buffer)
!$acc loop collapse(3)
do k = k_low, k_high
  do j = j_low, j_high
    do i = start_indx, end_indx
      n = n_old + (i - start_indx) + (j - j_low) * i_count + &
          (k - k_low) * j_count * i_count
      rho_buffer(n)   = soln%sblock(blck)%rho(i,j,k)
      vel_buffer(:,n) = soln%sblock(blck)%vel(:,i,j,k)
      p_buffer(n)     = soln%sblock(blck)%p(i,j,k)
      temp_buffer(n)  = soln%sblock(blck)%temp(i,j,k)
    end do
  end do
end do
!$acc end parallel
n = n_old + i_count * j_count * k_count
!$acc update host(rho_buffer(n_old:n-1))   async(1)
!$acc update host(vel_buffer(:,n_old:n-1)) async(2)
!$acc update host(p_buffer(n_old:n-1))     async(3)
!$acc update host(temp_buffer(n_old:n-1))  async(4)
\end{lstlisting}

However, when updating the buffer arrays on either side (device or host), since the host only transfers the derived type arrays such as \texttt{soln\%sblock\%array} not the buffer arrays \texttt{array\_buffer}, there is an extra step on the host side to pack/unpack the buffer to/from the derived type array, which can be seen in Listing~\ref{derived_array}. This step may not be needed for some other codes but necessary for SENSEI, as SENSEI uses derived type arrays to store primitive variables. The step adds some overhead to the host side, which will be addressed in $V5$.

\begin{lstlisting}[caption={An extra step to pack/unpack data to the derived type array},label=derived_array, language=Fortran]
start_indx = 1 - n_ghost_cells(1)
end_indx   = interior_cells
do k = k_low, k_high
  do j = j_low, j_high
    do i = start_indx, end_indx
      soln%sblock(blck)%rho(i,j,k)   = rho_buffer(n)
      soln%sblock(blck)%vel(:,i,j,k) = vel_buffer(:,n)
      soln%sblock(blck)%p(i,j,k)     = p_buffer(n)
      soln%sblock(blck)%temp(i,j,k)  = temp_buffer(n)
      n = n + 1
    end do
  end do
end do
\end{lstlisting}

\paragraph{V2: Extrapolating to ghost cells on the GPU}
The $V1$ version executes the kernel of extrapolating to ghost cells on the CPU. However, leaving the extrapolation on the CPU may impede further performance improvement as this portion will be the performance bottleneck for the GPU code. Therefore, $V2$ moves the kernel of extrapolating to ghost cells to the GPU. When passing an \texttt{intent(out)} reshaped array which is located in non-contiguous memory locations to a procedure call, the PGI compiler creates a temporary array that can be passed into the subroutine. The temporary array can reduce the performance significantly and poses a threat of cache contention if it is shared among CUDA threads. In fact, whether to support passing slices of array to a procedure call is a discussion for the NVIDIA PGI compiler group internally. To resolve this issue, manually created private temporary arrays are used to enable the GPU to parallelize the extrapolation kernel. An example of how the extrapolation works in SENSEI can be found in Listing~\ref{gc_extrapolation}. The \texttt{data present} directive notifies the compiler that the needed data are located in the GPU memory, the \texttt{data copyin} directive copies in the boundary information to the GPU, and the \texttt{parallel loop} directives parallelize the boundary loop iterations. The subroutine set\_bc is a device routine which is called in the parallel region. It is difficult for the compiler to automatically know whether there are loops inside the routine, and whether there are dependencies among the loop iterations in the parallel region. The use of \texttt{!\$acc routine seq} directive in set\_bc informs the compiler such information. After using the temporary arrays such as $rho$ and $vel$, each CUDA thread needs to have a copy of the arrays, which occupies a lot of SM registers and thus reduces the concurrency. As can been seen, these temporary arrays are used to store the data in the derived type in the beginning. Then they are used as arguments when invoking the set\_bc subroutine. Finally the extrapolated data are copied back to the ghost cells in the original derived type \texttt{soln}.

\begin{lstlisting}[caption={Using temporary array to do the ghost cell data extrapolation},label=gc_extrapolation, language=Fortran]
!$acc data present(soln, soln%rho, soln%vel, soln%p, &
!$acc              soln%temp, soln%molecular_weight, &
!$acc              grid%grid_vars%volume,            &
!$acc              grid%grid_vars%xi_n, grid, grid%grid_vars) &
!$acc      copyin(bound, bclow, bchigh, n_mmtm)

!$acc parallel
!$acc loop independent
do k = bound%indx_min(3),bound%indx_max(3)
  !$acc loop independent vector private(rho, vel, p, temp, vol)
  do j = bound%indx_min(2),bound%indx_max(2)
    rho(1:length)           = soln%rho(high+1:low:order,j,k)
    vel(1:n_mmtm,1:length)  = soln%vel(:,high+1:low:order,j,k)
    p(1:length)             = soln%p(high+1:low:order,j,k)
    temp(1:length)          = soln%temp(high+1:low:order,j,k)

    vol       = grid%grid_vars%volume(high:low:order,j,k)
    call set_bc(bound%bc_label,                               &
                rho,                                          &
                vel,                                          &
                p,                                            &
                temp,                                         &
                molweight,                                    &
                vol,                                          &
                grid%grid_vars%xi_n(:,i,j,k),                 &
                bclow,                                        &
                bchigh,                                       &
                n_mmtm)

    soln%rho(high+1:low:order,j,k)          = rho(1:length)
    soln%vel(1:n_mmtm,high+1:low:order,j,k) = vel(:,1:length)
    soln%p(high+1:low:order,j,k)            = p(1:length)
    soln%temp(high+1:low:order,j,k)         = temp(1:length)
  end do
end do
!$acc end parallel
!$acc end data

\end{lstlisting}

\paragraph{V3: Removal of Temporary Variables}
Either the automatic or the manual use of temporary arrays in $V2$ can greatly deteriorate the GPU performance. Instead of passing array slices to a subroutine, the entire array was passed with the indicies of the desired slice as shown in Listing~\ref{removing_temp}, which avoids the use of temporary arrays. This method requires many subroutines to be modified in SENSEI. However, it saves the use of shared resources and improves the concurrency.

\begin{lstlisting}[caption={Passing derived type data and index range},label=removing_temp, language=Fortran]
!$acc data present(soln, soln%rho, soln%vel, soln%p, &
!$acc              soln%temp, soln%molecular_weight, &
!$acc              grid%grid_vars%volume,            &
!$acc              grid%grid_vars%xi_n, grid, grid%grid_vars) &
!$acc      copyin(bound, bclow, bchigh, n_mmtm)

!$acc parallel
!$acc loop independent
do k = bound%indx_min(3),bound%indx_max(3)
  !$acc loop independent vector
  do j = bound%indx_min(2),bound%indx_max(2)
    call set_bc(bound%bc_label,               &
                grid,                         &
                soln,                         &
                soln%rho,                     &
                soln%vel,                     &
                soln%p,                       &
                soln%temp,                    &
                molweight,                    &
                grid%grid_vars%volume,        &
                grid%grid_vars%xi_n(:,i,j,k), &
                bclow,                        &
                bchigh,                       &
                j,                            &
                k,                            &
                n_mmtm,                       &
                boundary_lbl,                 &
                normal_lbl)

  end do
end do
!$acc end parallel
!$acc end data

\end{lstlisting}

\paragraph{V4: Splitting flux calculation and limiter calculation}

For cases which require the use of limiters, the CPU calculates the left and right limiters on a face once, as the next loop iteration can reuse two limiter values without computing them again, which can be seen in Eq.~\ref{flux}.

\begin{align}
\label{flux}
\vec{Q}_{i+1/2}^L = & \vec{Q}_i + \frac{\epsilon}{4} [(1 - \kappa) \Psi_{i-1/2}^{+} (\vec{Q}_{i} - \vec{Q}_{i-1}) + (1 + \kappa) \Psi_{i+1/2}^{-} (\vec{Q}_{i+1} - \vec{Q}_{i})] \\
\vec{Q}_{i+1/2}^R = & \vec{Q}_{i+1} - \frac{\epsilon}{4} [(1 + \kappa) \Psi_{i+1/2}^{+} (\vec{Q}_{i+1} - \vec{Q}_{i}) + (1 - \kappa) \Psi_{i+3/2}^{-} (\vec{Q}_{i+2} - \vec{Q}_{i+1})]
\end{align}
where $\epsilon$ and $\kappa$ are MUSCL extrapolation parameters, $\Psi$ are limiter function values. $L$ and $R$ denote the left and right states, respectively.

After porting the code to the GPU, since SENSEI calculates the limiters locally for each solution state (in $V0$ through $V3$), the limiter cannot be reused as different CUDA threads have their own copies of four limiter values, otherwise thread contention may occur. To fix this issue, the total cost of the limiter calculation on the GPU is twice of that on the CPU. Also, storing the limiter locally requires the limiter calculation and flux extrapolation to be together, which is highly compute intensive. $V4$ uses global arrays to store these limiters so that the flux calculation and limiter calculation can be separated, which is given in listing~\ref{split_muscl_limiter}. This approach will leave more room for kernel concurrency and asynchronization and also avoid thread contention.

\begin{lstlisting}[caption={Splitting MUSCL extrapolation and limiter calculation},label=split_muscl_limiter, language=Fortran]
  ! xi limiter
  !$acc parallel
  !$acc loop independent collapse(3)
  do k = 1, k_cells
    do j = 1, j_cells

      do i = 1, imax-1
        call limiter_subroutine_x( sblock, gblock, i, j, k,  &
                                   sblock%limiter_xi%left,   &
                                   sblock%limiter_xi%right )
      end do
    end do
  end do
  !$acc end parallel

  ! xi flux
  !$acc parallel
  !$acc loop independent collapse(3) private(qL, qR)
  do k = 1, k_cells
    do j = 1, j_cells
      do i = 2, imax-1

        call muscl_extrapolation_xi( sblock, i, j, k, &
                   sblock%limiter_xi%left(1:neq,i-1,j,k),  &
                   sblock%limiter_xi%left(1:neq,i,j,k),    &
                   sblock%limiter_xi%right(1:neq,i,j,k),   &
                   sblock%limiter_xi%right(1:neq,i+1,j,k), &
                   qL, qR )

        call flux_function(qL, qR, &
                           gblock%grid_vars%xi_n(:,i,j,k), &
                           sblock%xi_flux(1:neq,i,j,k))

      end do

    end do
  end do
  !$acc end parallel

\end{lstlisting}

\paragraph{V5: Derived type for connected boundaries on the GPU}
The previous versions update the connected boundaries between the host and the device through using local dynamic arrays. Therefore, it is worthwhile to investigate the effect of using global derived type arrays to store the connected boundary data. It removes the extra data copies on the host side mentioned in $V1$. An example of using the global derived type is given in Listing~\ref{derived_boundary}. If there is no communication required among different CPU processors, the MPI functions are not called.

\begin{lstlisting}[caption={Derived type for connected boundary data},label=derived_boundary, language=Fortran]
!$acc update host(grid%gblock(blck)%bcs_acc(nc)%rho_send(   &
!$acc             1:idx_max_nbor(1)-idx_min_nbor(1)+1,      &
!$acc             1:idx_max_nbor(2)-idx_min_nbor(2)+1,      &
!$acc             1:idx_max_nbor(3)-idx_min_nbor(3)+1))

! SEND and RECV derived type boundary data
call MPI_IRECV( grid%gblock(blck)%bcs_acc(nc)%rho_recv,     &
                scalar_count, MPI_DOUBLE_PRECISION,         &
                bound%bound_nbor%process_id, RHO_TAG,       &
                world_comm, req(req_count+1), ierr )

call MPI_ISEND( grid%gblock(blck)%bcs_acc(nc)%rho_send,     &
                scalar_count, MPI_DOUBLE_PRECISION,         &
                bound%bound_nbor%process_id, RHO_TAG,       &
                world_comm, req(req_count+5), ierr )

call MPI_WAITALL(req_count, req(1:req_count),               &
                 stat(:,1:req_count), ierr)

!$acc update device(grid%gblock(blck)%bcs_acc(nc)%rho_recv( &
!$acc               buff_size_self(1)*buff_size_self(2)*    &
!$acc               buff_size_self(3)))

\end{lstlisting}

\paragraph{V6: Change of blocking call locations}

\begin{figure}[H]
	\centering
	\includegraphics[width=.7\textwidth]{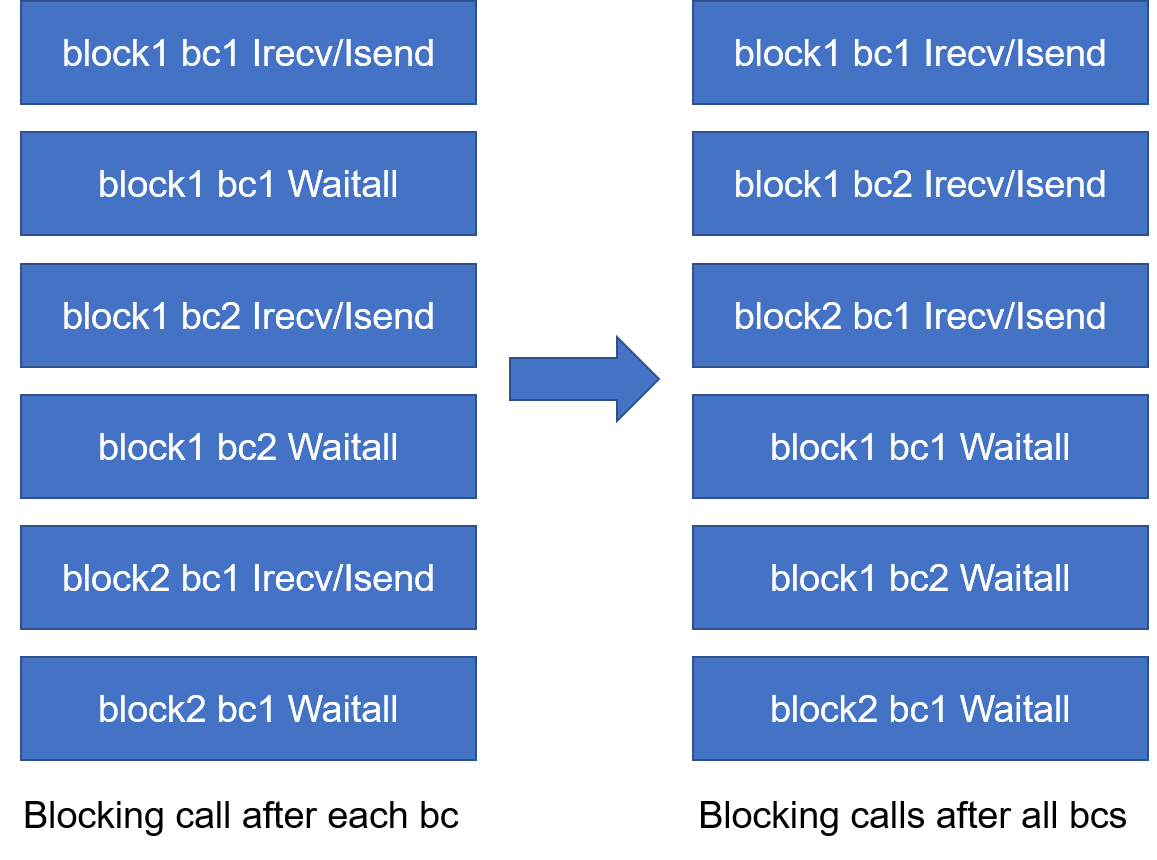}
	\caption{Change of blocking call position}
	\label{change_block_position}
\end{figure}

Since SENSEI is a multi-block CFD code, a processor may hold multiple blocks and many connected boundaries. Using MPI non-blocking routines, there should be a place to execute the blocking call such as MPI\_WAIT to complete the communications. Each Isend/Irecv call needs one MPI\_WAIT, or multiple MPI\_WAIT can be wrapped up into one MPI\_WAITALL. The previous versions block the MPI\_WAITALL call for every decomposed block. A newer way of achieving the function is moving the MPI\_WAITALL calls to a new loop, so that these MPI\_WAITALL calls are executed after all Isend \& Irecv are executed. An example is given in Fig.~\ref{change_block_position}. In this example, there are two blocks, each having two connected boundaries. However, $V6$ only improves the performance when multiple connected boundaries exist.

For platforms in which the asynchronous progression is supported completely (from both the software and hardware sides), this optimization may work much better. However, for common platforms in which the asynchronous progression is not supported fully, OpenMP may need to be used to promote the asynchronous progression~\cite{jiayin2006overlapping,vaidyanathan2015improving,lu2015mpi+,denis2016mpi,castillo2019optimizing}. Full asynchronous progression is a very complicated issue and is not covered in this paper. This paper will only apply MPI+OpenACC to accelerate the CFD code.

\paragraph{V7: Boundary flux optimization}

In SENSEI, the fluxes for the wall and farfield boundaries need to be overwritten to get more accurate estimate for the solution. These overwritten flux calculations are done after the boundary enforcement. For these two kinds of fluxes, the previous versions do not compute them very efficiently. A lot of temporary variables are allocated for each thread, which deteriorates the concurrency of using OpenACC, as registers are limited. The principle of this optimization is similar to that in $V3$. An example of the optimization is given in Listing~\ref{boundary_flux}. 

\begin{lstlisting}[caption={Optimization of the overwritten boundary flux kernel},label=boundary_flux, language=Fortran]
! V0 ~ V6

!$acc parallel copyin(i, bound) async(1)
!$acc loop independent
do k = bound%indx_min(3), bound%indx_max(3)
  !$acc loop independent vector private(      &
  !$acc      soln_L2, soln_L1, soln_R1,       &
  !$acc      soln_R2, qL, qR, modf,           &
  !$acc      lim_L2, lim_L1, lim_R1, lim_R2,  &
  !$acc      vel_xi, rho_xi, p_xi, temp_xi)
  do j = bound%indx_min(2), bound%indx_max(2)

! V7

!$acc parallel copyin(i, bound) async(1)
!$acc loop independent
do k = bound%indx_min(3), bound%indx_max(3)
  !$acc loop independent vector private(      &
  !$acc      qL, qR, modf)
  do j = bound%indx_min(2), bound%indx_max(2)

\end{lstlisting}

\paragraph{V8: Asynchronicity improvement}
Kernels from different streams can be overlapped so that the performance can be improved. The version is exactly the same as that in $V7$ but the environment variable "PGI\_ACC\_SYNCHRONOUS" is set to 0 when executing SENSEI, that is, asynchronization among some independent kernels is promoted. The \texttt{!\$acc wait} directive makes the host wait until asynchronous accelerator activities finish, i.e., it is the synchronization on the host side.

\paragraph{V9: Removal of implicit data copies between the host and device}
The last performance optimization is essentially manual tuning work. It requires the user to modify the code through profiling. The compiler sometimes does not know what variables are to be updated between the host and the device, so for the reason of safety the compiler may update variables frequently, which may be unnecessary. Different architectures and compilers may deal with the update differently, therefore the user can optimize it based on the profiler outputs. The compiler may transfer some scalar variables, arrays with small size and even derived type data in every iteration, but they only need to be copied once. There are multiple places in SENSEI where the PGI compiler makes unnecessary copies. These extra unnecessary data transfers are usually small in size and deteriorate the memory throughput. The effect of these copies can be significant for small size problems. However, for compute-intensive computations, this optimization may not be very useful. This performance optimization is only applied for the P100 GPU and V100 GPU, with the newer version of PGI compiler. Running with $V9$ on the C2075 returns some linker errors due to the old PGI compiler. 

\paragraph{V10: GPUDirect}
GPUDirect is an umbrella word for several GPU communication acceleration technologies. It provides high bandwidth and low latency communication between NVIDIA GPUs. There are three levels of GPUDirect~\cite{GPUDirect}. The first level is GPUDirect Shared Access, introduced with CUDA 3.1. This feature avoids an unnecessary memory copy within host memory between the intermediate pinned buffers of the CUDA driver and the network fabric buffer. The second level is GPUDirect Peer-to-Peer transfer (P2P transfer) and Peer-to-Peer memory access (P2P memory access), introduced with CUDA 4.0. This P2P memory access allows buffers to be copied directly between two GPUs on the same node. The last is GPU RDMA (Remote Direct Memory Access), with which buffers can be sent from the GPU memory to a network adapter without staging through host memory. The last feature is not supported on NewRiver as it pertains to specific versions of the drivers (from NVIDIA and Mellanox for the GPU and the Infiniband, respectively) which are not installed (other dependencies exist, particularly parallel filesystems). Although GPU RDMA is not available, the other aspects of GPUDirect can be utilized to further improve the scaling performance on multiple GPUs.

\section{Solution and Scaling Performance}

\subsection{Supersonic Flow Through a 2D Inlet}
The first test case is a simplified 2D 30 degree supersonic inlet, which has only one parent block without having connected boundaries. The inflow conditions are given in Table~\ref{inlet_inflow}. There are multiple levels of grid for strong and weak scaling analysis, of which the total amount of cells range from 50k to 7 million. The parallel solution and the serial solution have been compared from the beginning to the converged state during the iterations, and the relative errors for all the primitive variables based on the inflow boundary values is within round-off error range ($10^{-12}$).

\begin{table}[H]
	\caption{Inlet case inflow boundary conditions}
	\centering
	\begin{tabular}{cc}
		\hline
		Mach number& 4.0\\
		Pressure& 12270 Pa\\
		Temperature& 217 K\\
		\hline
	\end{tabular}
	\label{inlet_inflow}
\end{table}

\begin{figure}[H]
	\centering 
	\subfigure[A coarse (52x16) grid for the 2D inlet flow]{ 
		\label{inlet-grid}
		\includegraphics[width=.45\textwidth,trim=7 7 7 7,clip]{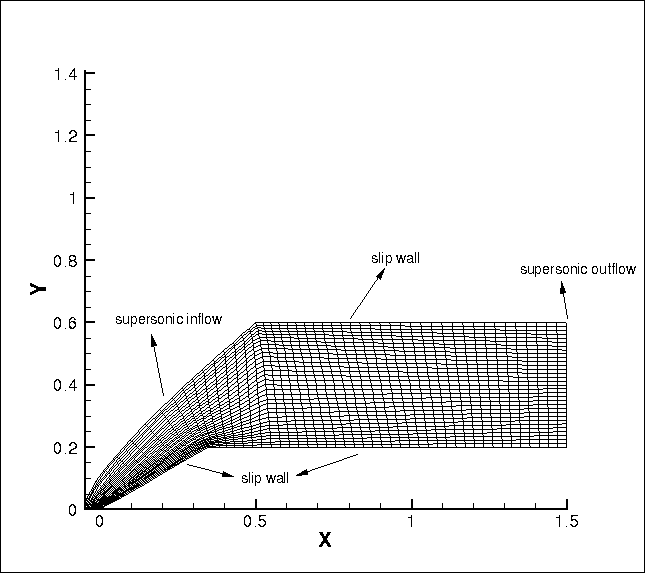} 
	}
	\subfigure[A domain decomposition for the 2D inlet flow (using 16 GPUs)]{ 
		\label{inlet-id}
		\includegraphics[width=.45\textwidth,trim=7 7 7 7,clip]{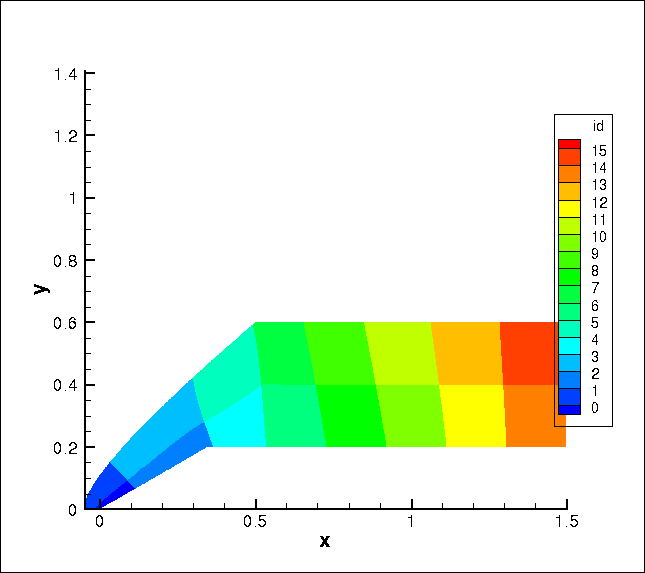} 
	} 
	\caption{2D Euler supersonic inlet} 
	\label{Inlet-grid-id}
\end{figure}

 A very coarse level of grid for the 2D inlet flow is shown in Fig.~\ref{inlet-grid}. The decomposition of using 16 GPUs (which is the highest number of GPUs available) on a 416x128 grid is shown in Fig.~\ref{inlet-id}. The decomposition is 2D, creating multiple connected boundaries between processors. Ghost cells on the face of connected boundaries are used to exchange data between neighboring processors. The device needs to communicate with the host if multiple processors are used.

\begin{figure}[H]
	\centering
	\includegraphics[width=.6\textwidth,trim=7 7 7 7,clip]{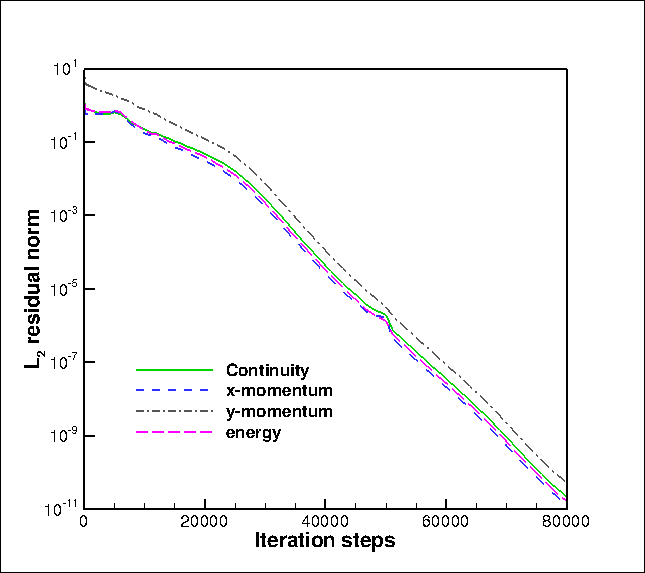}
	\caption{The relative iterative residual history for the inlet case}
	\label{inlet-resid}
\end{figure}

\begin{figure}[H]
	\centering 
	\subfigure[The Mach number and streamlines for the 2D inlet Euler flow]{ 
		\label{inlet_Ma}
		\includegraphics[width=.45\textwidth,trim=7 7 7 7,clip]{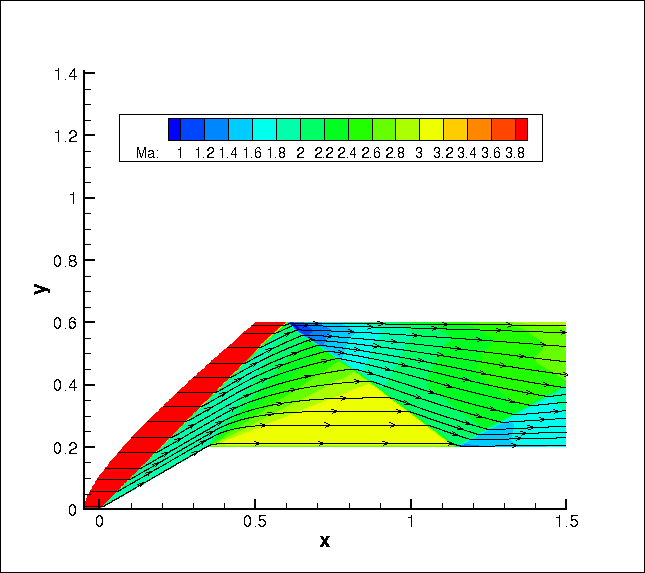} 
	} 
	\subfigure[The density solution for the 2D inlet Euler flow]{ 
		\label{inlet_rho_euler}
		\includegraphics[width=.45\textwidth,trim=7 7 7 7,clip]{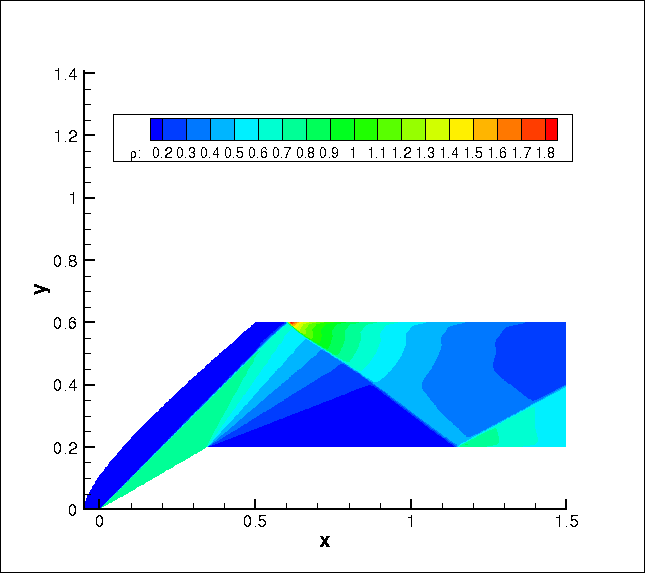} 
	} 
	\caption{2D Euler supersonic inlet} 
	\label{Inlet-Mach-rho}
\end{figure}

 The relative residual $L_2$ norm history is shown in Fig.~\ref{inlet-resid}. It can be seen that the iterative errors have been driven down small enough for all the primitive variables when converged. The Mach number and density solutions are shown in Fig.~\ref{Inlet-Mach-rho}. There are multiple flow deflections when the flow goes through the reflected oblique shocks.

\begin{figure}[H]
	\centering
	\includegraphics[width=.7\textwidth]{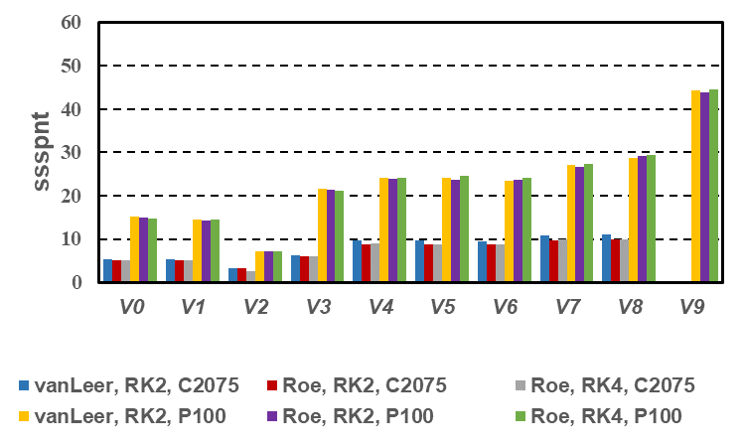}
	\caption{Performance comparison for the 2D inlet Euler flow}
	\label{inlet-versions}
\end{figure}

Fig.~\ref{inlet-versions} shows the performance comparison of different optimizations using different flux options on different platforms. The grid size used in Fig.~\ref{inlet-versions} is 416$\times$128. The goal of making such a comparison is to investigate the effect of using various flux options, time marching schemes and various generation GPUs when applying the optimizations introduced earlier in this paper. For such a small problem which does not have any connected boundary conditions, a single P100 GPU is about 3 times faster than the a single C2075 GPU. We expect that the speedup would be higher if the problem size was larger. Another observation is that using the Roe flux is slightly slower than using the van Leer flux, which is reasonable as the Roe flux is a bit more expensive than the van Leer flux. It should be kept in mind that the ssspnt metric does not take the number of double precision operations for each step into account so ssspnt is not equivalent to GFLOPS. Also, the speed of RK2 and RK4 is comparable, so this paper will stick to the use of RK2 unless otherwise specified.

If comparing the performance of different versions in Fig.~\ref{inlet-versions}, there are two performance leaps including from $V2$ to $V3$ and from $V8$ to $V9$. Since the extrapolation to ghost cells on the GPU runs inefficiently in $V2$ due to the low compute utilization, removing the use of temporary arrays in the parallel regions reduces the overhead from CUDA threads. More concurrency in the code can therefore be utilized by the GPU. From $V8$ to $V9$, since the problem size is small (the compute fraction is not very high), removing unnecessary data movement improves the overall performance by more than 52\%. For larger problems, the performance gain is not that significant, as we will show later. In the meantime, there is a gradual performance improvement from $V3$ to $V4$ and $V6$ to $V8$. These optimizations should not be overlooked as the issues related to the optimizations will eventually become bottlenecks. Since this case does not have connected boundaries, there is no obvious performance change from $V4$ to $V6$. It should be mentioned that the performance optimizations proposed earlier are not for only a specific case, but for general cases with multiple blocks and connected boundaries.

\begin{figure}[H]
	\centering 
	\subfigure[Strong scaling]{ 
		\label{inlet-strong}
		\includegraphics[width=.45\textwidth,trim=7 7 0 7,clip]{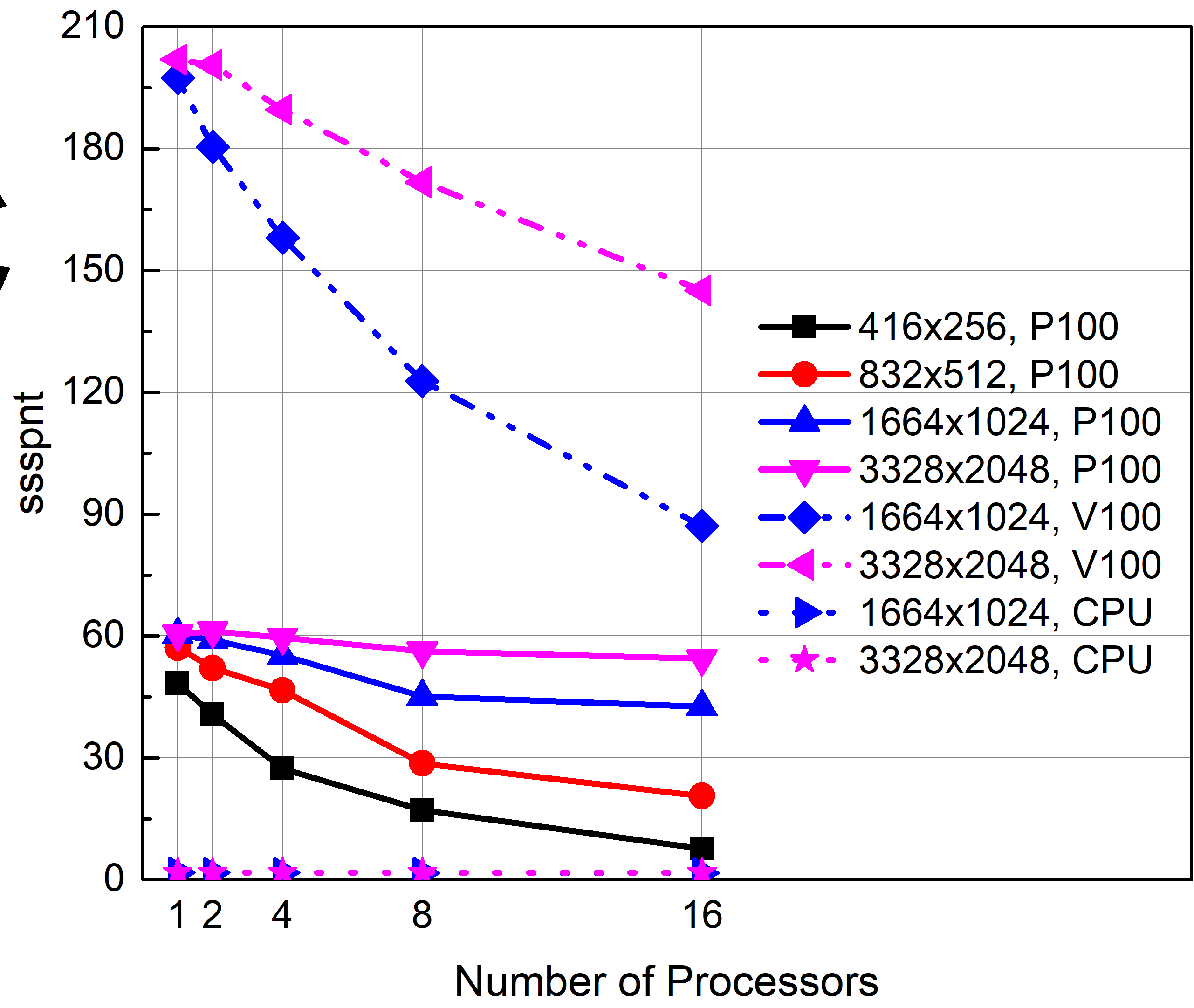} 
	} 
	\subfigure[Weak scaling]{ 
		\label{inlet-weak}
		\includegraphics[width=.45\textwidth,trim=7 7 0 7,clip]{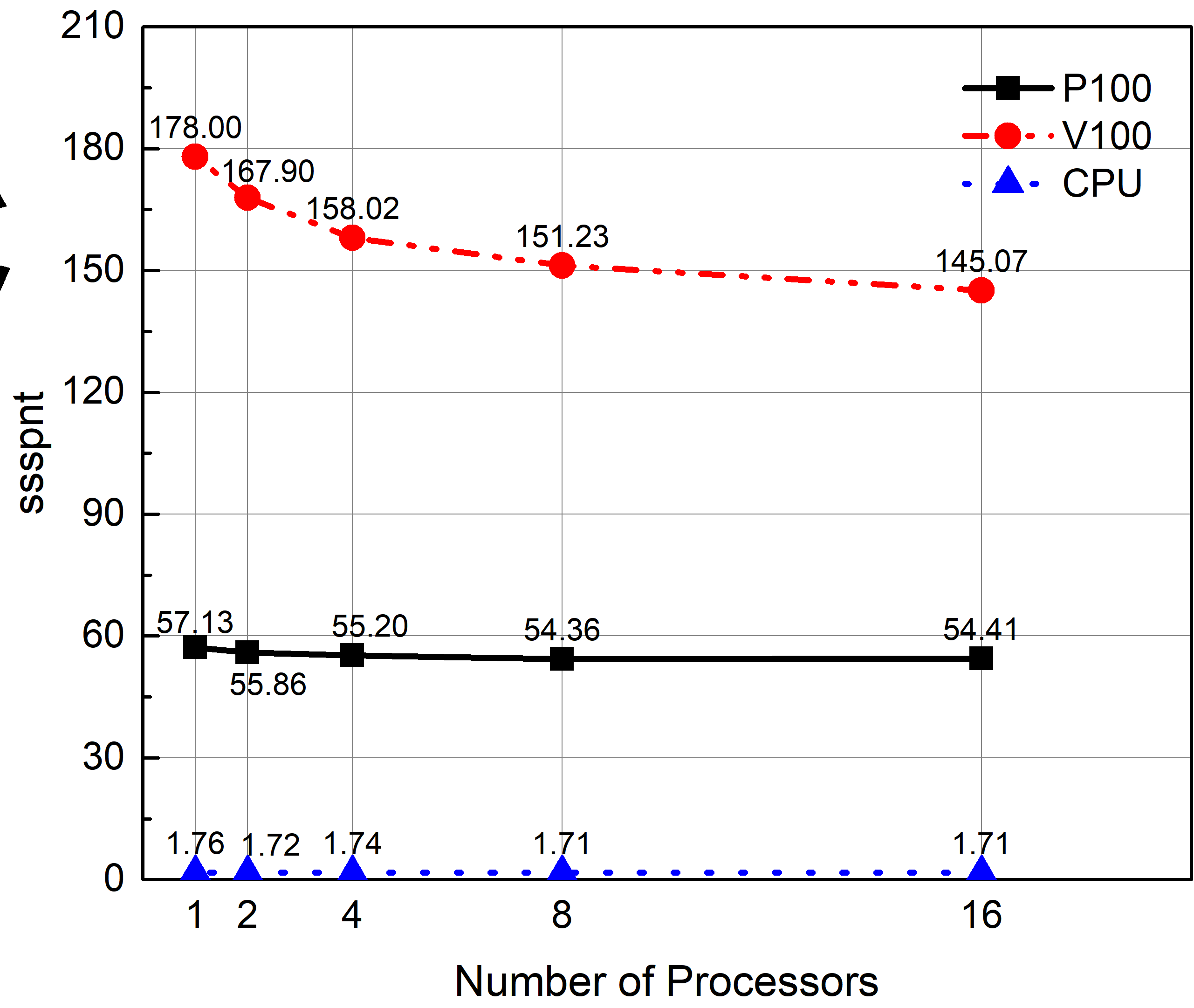} 
	} 
	\caption{The scaling performance for the 2D inlet case} 
	\label{Inlet}
\end{figure}

Fig.~\ref{inlet-strong} and Fig.~\ref{inlet-weak} show the strong and weak scaling performance for the 2D inlet Euler flow, respectively. The CPU scaling performance is also given for reference. A single P100 GPU is more than 32$\times$ faster than a single CPU, on a grid level of 416x256, which displays the compute power of the GPU. The strong scaling efficiency decays quickly for small problem sizes but not for the largest problem size in Fig.~\ref{inlet-strong}. The parallel efficiency using 16 P100 GPUs on the 3328$\times$2048 grid is still kept higher than 90\%. While for the weak scaling, the parallel efficiency is higher (95.2\% above) than the strong scaling efficiency, as there is more work to saturate the GPU. The V100 GPU shows higher speedups but lower efficiency, because the V100 GPU needs more computational work as it is faster. The boundary connections for this inlet flow case after the domain decomposition are not complicated, which is one important reason why the performance is very good.

\subsection{2D Subsonic Flow past a NACA 0012 Airfoil}

The second test case in this paper is the 2D subsonic flow ($M_\infty=0.25$) past a NACA 0012 airfoil, at an angle of attack of 5 degrees. The flow field for all the simulation runs of this case is initialized using the farfield boundary conditions which are given in Table~\ref{airfoil_farfield}. This case will be solved by both the Euler and laminar NS solvers in SENSEI.

\begin{table}[H]
	\caption{NACA 0012 airfoil farfield boundary conditions}
	\centering
	\begin{tabular}{cc}
		\hline
		Mach number& 0.25\\
		Static pressure& 84307 Pa\\
		Temperature& 300 K\\
		Angle of attack, $\alpha$& 5 degrees\\
		\hline
	\end{tabular}
	\label{airfoil_farfield}
\end{table}

\begin{figure}[H]
	\centering 
	\subfigure[A coarse (128x48) grid for the flow past a NACA 0012 airfoil]{ 
		\label{naca0012-grid}
		\includegraphics[width=.45\textwidth,trim=7 7 7 7,clip]{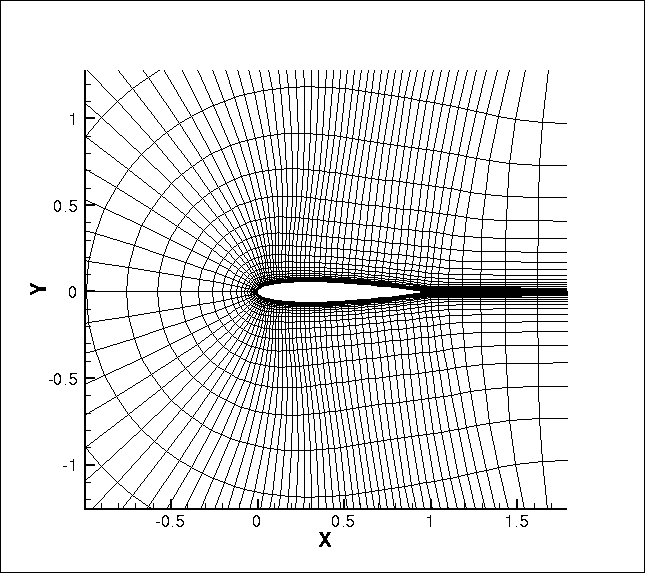} 
	} 
	\subfigure[The domain decomposition for the airfoil case (using 16 GPUs)]{ 
		\label{naca0012-id}
		\includegraphics[width=.45\textwidth,trim=7 7 7 7,clip]{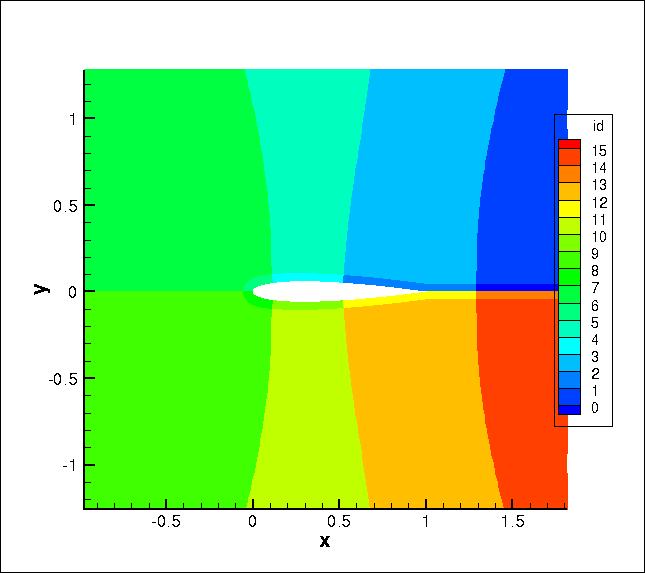} 
	} 
	\caption{2D NS NACA 0012 airfoil} 
	\label{naca0012-grid-id}
\end{figure}

Although the airfoil case contains only one parent block, the grid is a C-grid, which means that the only one block connects to itself on a face through a connected boundary, which makes the airfoil case different from the 2D inlet flow case. One coarse grid of this airfoil case is shown in Fig.~\ref{naca0012-grid}. For the scaling analysis, the grid size ranges from 400k to 6 million. Also, the domain decomposition of using 16 GPUs is shown in Fig.~\ref{naca0012-id}. Near the airfoil surface, the grid is refined locally so processors near the wall take smaller blocks, but the load is balanced.

\begin{figure}[H]
	\centering 
	\subfigure[The relative iterative residual $L_2$ norm history for the laminar NS subsonic flow past a NACA airfoil]{ 
		\label{naca0012-resid}
		\includegraphics[width=.45\textwidth,trim=7 7 7 7,clip]{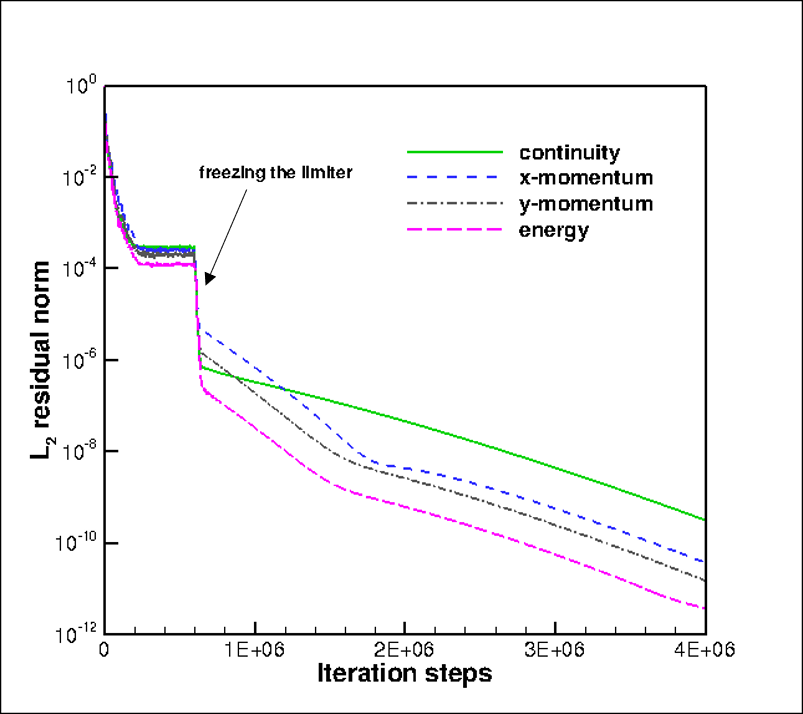} 
	} 
	\subfigure[The pressure coefficient contour for the laminar NS subsonic flow past a NACA airfoil]{ 
		\label{naca0012_cp_NS}
		\includegraphics[width=.45\textwidth,trim=7 7 7 7,clip]{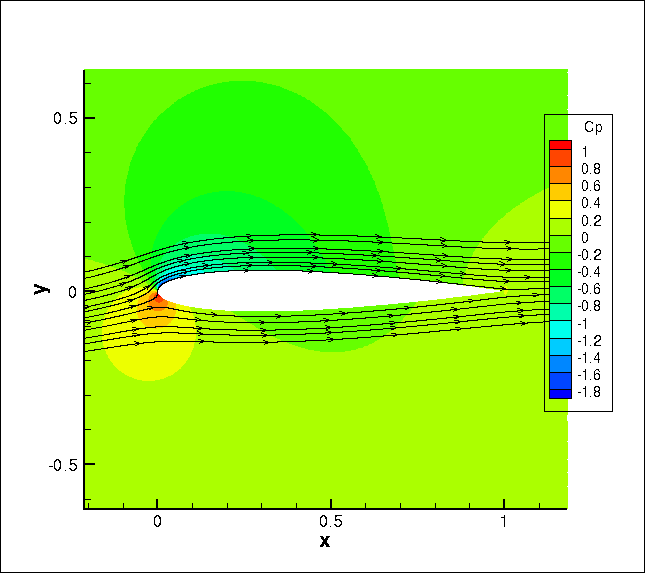} 
	} 
	\caption{2D laminar NS NACA 0012 airfoil} 
	\label{naca0012-iter-soln}
\end{figure}

Fig.~\ref{naca0012-resid} shows the relative iterative residual $L_2$ norm history for the laminar NS subsonic flow past a NACA 0012 airfoil. This case requires the most iteration steps to be converged among all the test cases considered. Leveraging the compute power of the GPU saves a lot of time. To enable the iterative residual to further go down instead of oscillation, limiter freezing is adopted at around 600k steps. After freezing the limiter, the iterative residual norms continue to reduce smoothly. The iterative errors are driven down small enough to obtain the steady state solution. 

The parallel solution and the serial solution have been compared on coarse levels of grid and the relative errors for all the primitive variables based on the reference values are within round-off error range ($10^{-12}$). Fig.\ref{naca0012_cp_NS} shows the pressure coefficient solution and the streamlines for the laminar NS subsonic flow past the NACA 0012 airfoil.

\begin{figure}[H]
	\centering
	\includegraphics[width=.7\textwidth,trim=7 7 7 7,clip]{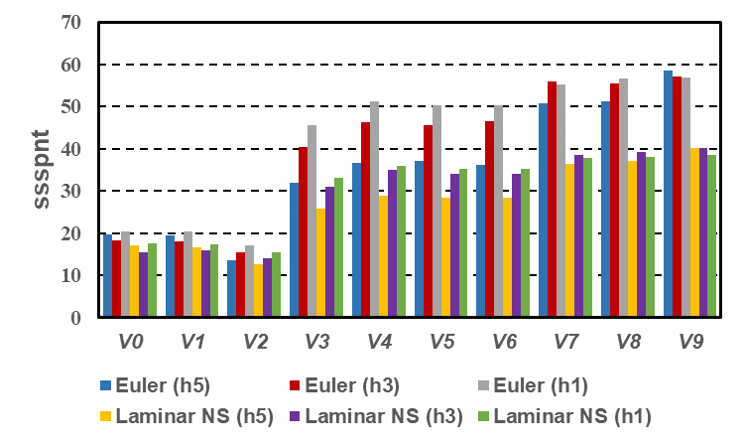}
	\caption{The performance of different versions for the NACA 0012 airfoil case (P100 GPU)}
	\label{airfoil-versions}
\end{figure}

Fig.~\ref{airfoil-versions} shows the comparison of different versions for the flow past a NACA 0012 airfoil using a single P100 GPU. Laminar NS has a smaller ssspnt (about 70\%) compared to using the Euler solver. From $V2$ to $V3$, the speedup is more than 2 times on different levels of grid, for both the Euler and laminar NS solver. To use globally allocated derived types to store the connected boundary data cannot improve the performance, which can be seen from the comparison of $V4$ and $V5$, if only using one processor, as there are no MPI communication calls. Although the airfoil case has a connected boundary, the data in the ghost cells for that boundary are filled directly through copying. This case only has one connected boundary, so there is no need to reorder the non-blocking MPI I\_send/I\_recv calls and the MPI\_Wait call. Similarly to the 2D inlet case, on coarse levels of grid, there is noticeable performance improvement if applying the optimization in $V9$. On fine levels of grid, the benefit is limited.

\begin{figure}[H]
	\centering
	\includegraphics[width=.7\textwidth,trim=7 7 7 7,clip]{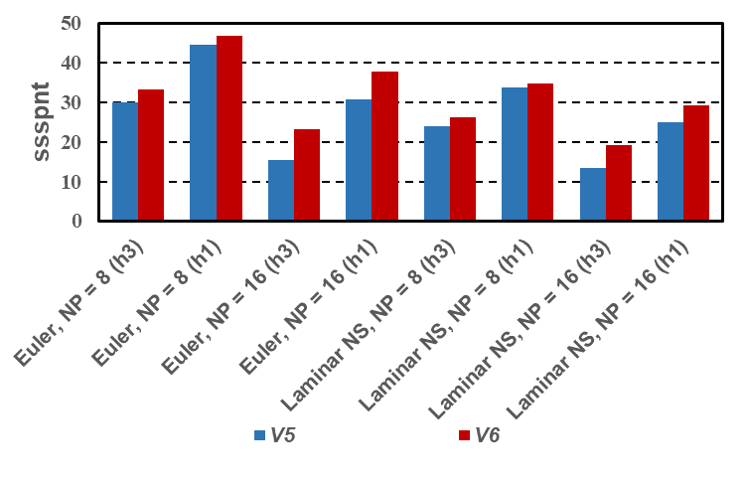}
	\caption{Performance comparison between $V5$ and $V6$ for the NACA 0012 airfoil case (P100 GPU)}
	\label{naca0012-V5V6}
\end{figure}

Since we cannot see any performance gain from $V5$ to $V6$ using single GPU, multiple GPUs are used to show the benefits. For all the runs shown in Fig.~\ref{naca0012-V5V6}, $V6$ (the red bars) outperforms $V5$ (the blue bars) by 4\% to 50\%, depending on the solver type, grid level and number of GPUs used. After applying multiple GPUs, multiple connected boundaries are created, which creates margin for the reordering of I\_send/I\_recv and Wait to work. Intrinsically, this ordering is to propel more asynchronous progression on the implementation side. The actual overlap degree still highly depends on the communication system, which is out of the scope of this paper. Readers who are interested in more overlap and better asynchronous progression may try the combination of MPI+OpenACC+OpenMP.

\begin{figure}[H]
	\centering 
	\subfigure[Strong scaling (Euler)]{ 
		\label{euler-airfoil-strong}
		\includegraphics[width=.45\textwidth,trim=7 7 0 7,clip]{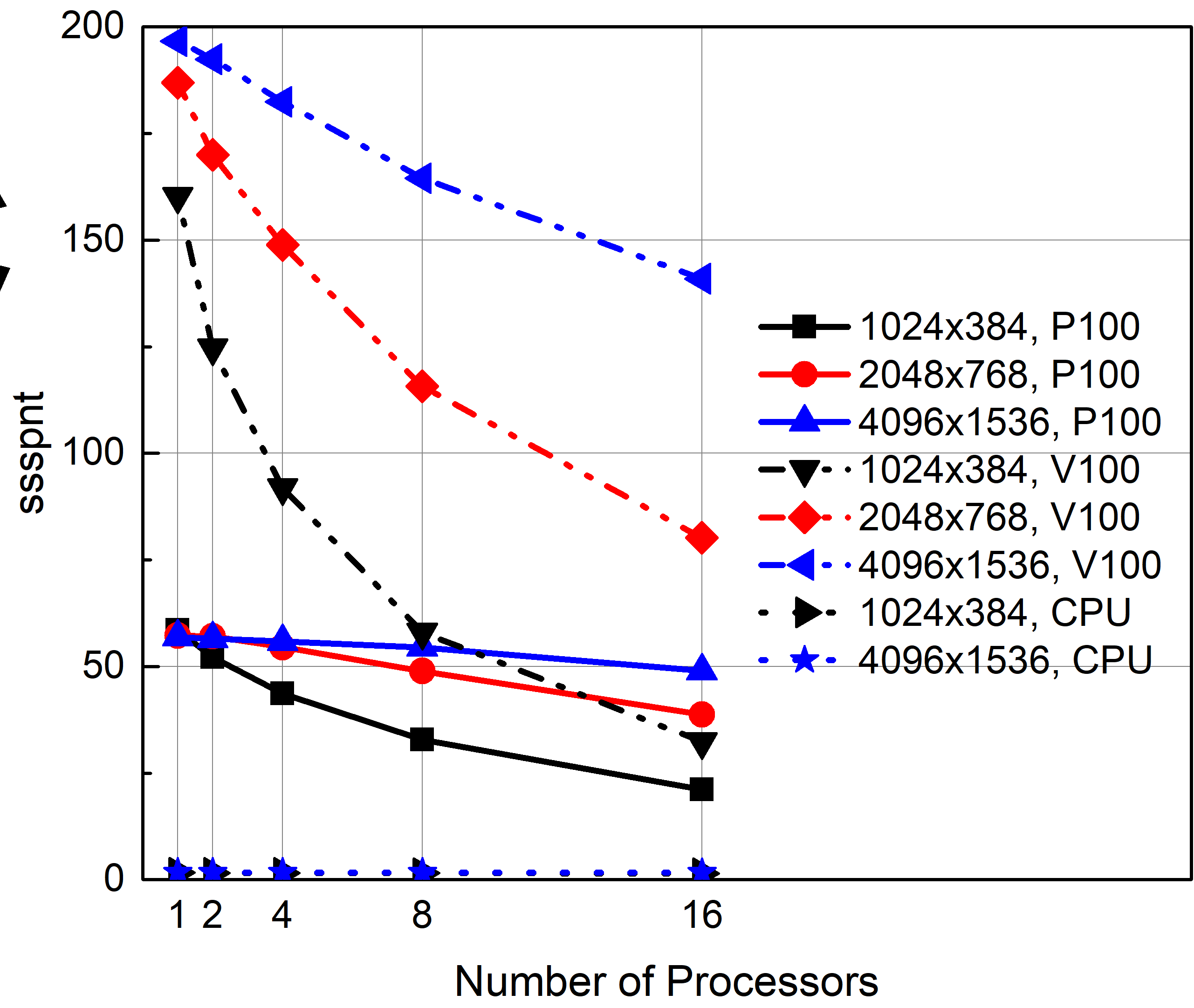} 
	} 
	\subfigure[Weak scaling (Euler)]{ 
		\label{euler-airfoil-weak}
		\includegraphics[width=.45\textwidth,trim=7 7 0 7,clip]{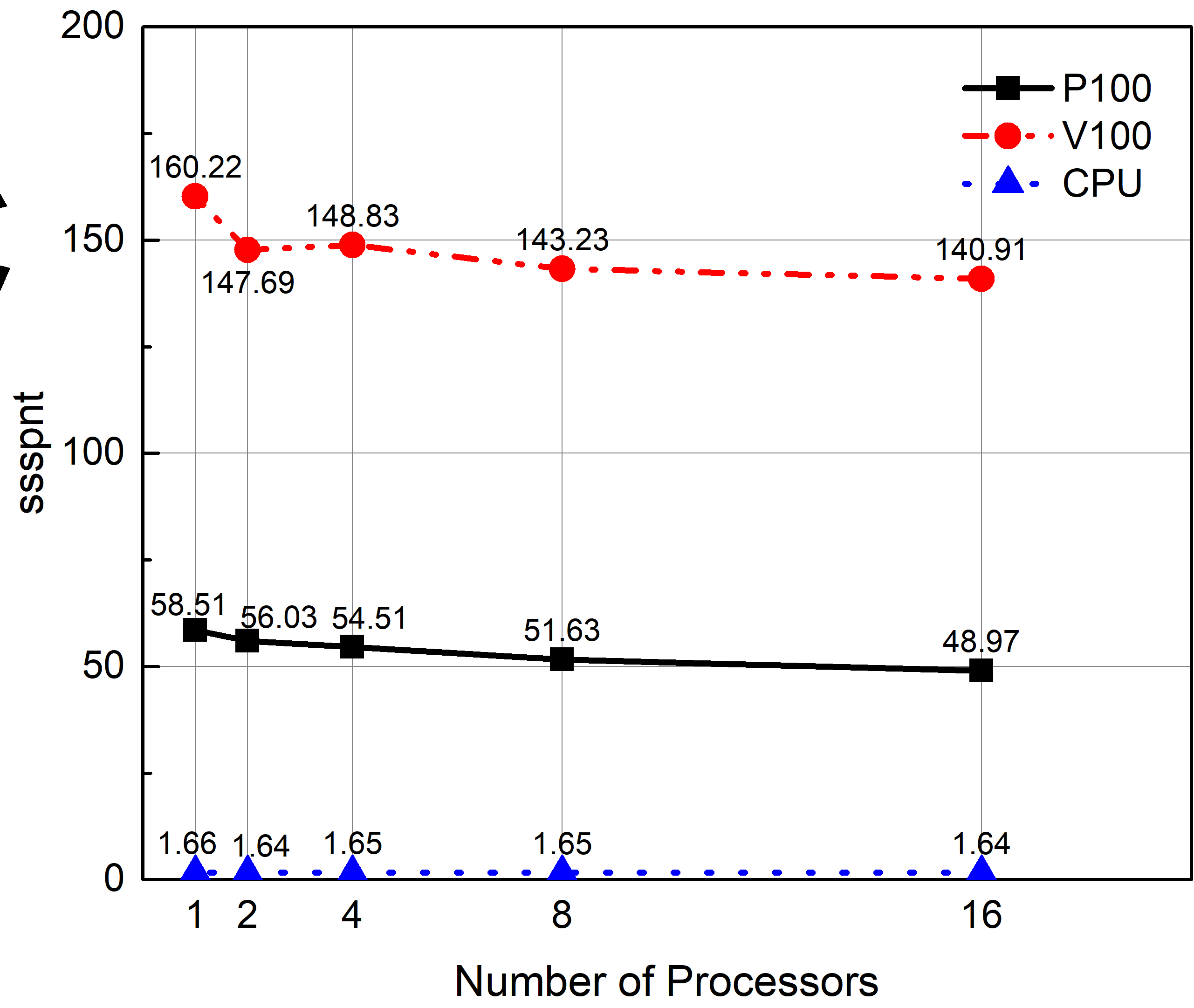} 
	} 
	\caption{The scaling performance for the 2D Euler flow past a NACA 0012 airfoil} 
	\label{Euler_airfoil}
\end{figure}

\begin{figure}[H]
	\centering 
	\subfigure[Strong scaling (laminar NS)]{ 
		\label{NS-airfoil-strong}
		\includegraphics[width=.45\textwidth,trim=7 7 0 7,clip]{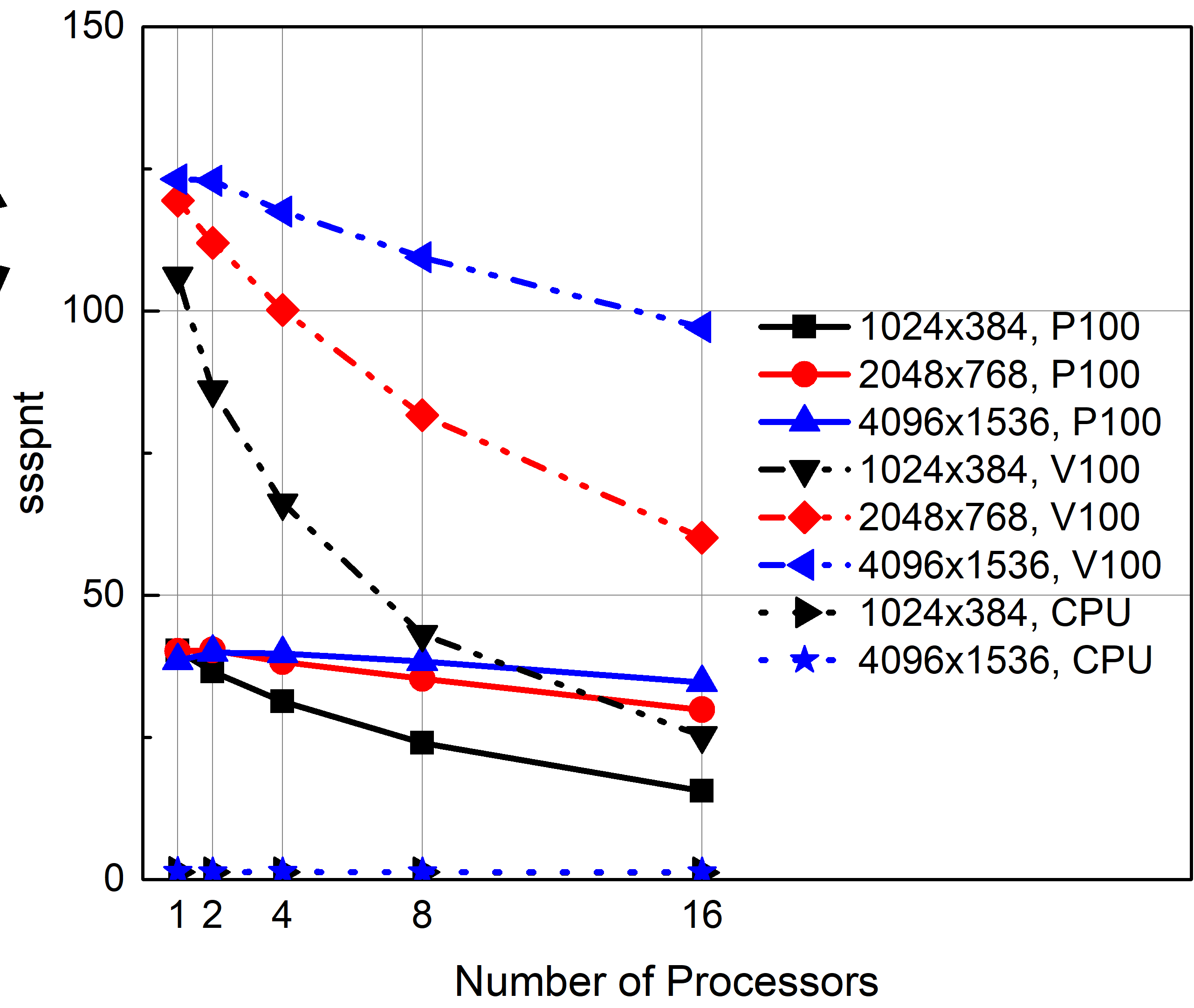} 
	} 
	\subfigure[Weak scaling (laminar NS)]{ 
		\label{NS-airfoil-weak}
		\includegraphics[width=.45\textwidth,trim=7 7 0 7,clip]{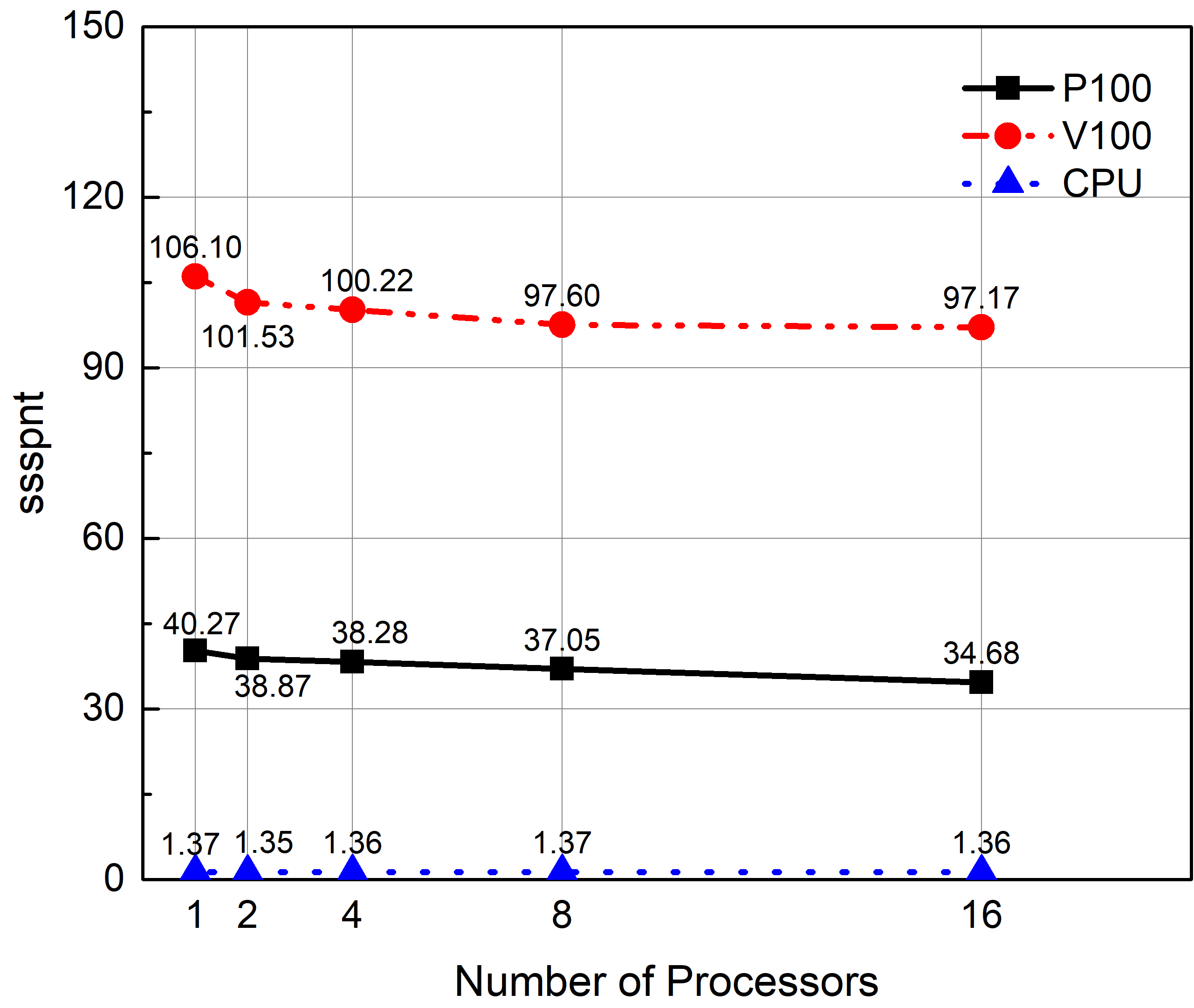} 
	} 
	\caption{The scaling performance for the 2D laminar NS flow past a NACA 0012 airfoil} 
	\label{NS_airfoil}
\end{figure}

Fig.~\ref{Euler_airfoil} and Fig.~\ref{NS_airfoil} show the strong and weak scaling performance of this subsonic flow past a NACA 0012 airfoil solved by the Euler and laminar NS solver on P100 and V100 GPUs, respectively. They show very similar behaviours with the only difference in the scales. Overall, the laminar ssspnt is about 0.7 of the Euler ssspnt using multiple GPUs. The strong parallel efficiency on the 4096$\times$1536 grid using 16 P100 GPUs for the Euler and laminar NS solver is about 87\% and 90\%, respectively. The weak scaling efficiency is generally higher as there is more work to do for the GPU. The efficiencies using V100 GPUs are lower than those using P100 GPUs, which indicates that faster GPUs may need more computational work to hold high efficiency.

\subsection{3D Transonic Flow Past an ONERA M6 Wing}

The final case tested in this paper is the 3D transonic flow ($M_\infty=0.839$) past an ONERA M6 wing, at an angle of attack of 3.06 degrees~\cite{mani1997assessment}. The flow field is initialized using the farfield boundary conditions which are given in Table~\ref{wing_farfield}. Both the Euler and laminar NS solvers in SENSEI are used to solve this problem. Different from the previous two 2D problems, this 3D case has 4 parent blocks with various sizes. Under some conditions (when using 2 and 4 processors in this paper), domain aggregation is needed to balance the load on different processors. This 3D wing case has a total grid size ranging from 300k to 5 million.

\begin{table}[H]
	\caption{ONERA M6 wing farfield boundary conditions}
	\centering
	\begin{tabular}{cc}
		\hline
		Mach number, $M_\infty$& 0.8395\\
		Temperature, $T_\infty$& 255.556 K\\
		Pressure, $p_\infty$& 315979.763 Pa\\
		Angle of attack, $\alpha$& 3.06 degrees\\
		\hline
	\end{tabular}
	\label{wing_farfield}
\end{table}

\begin{figure}[H]
	\centering 
	\subfigure[A grid for the ONERA M6 wing]{ 
		\label{m6onera_grid}
		\includegraphics[width=.45\textwidth,trim=7 7 7 7,clip]{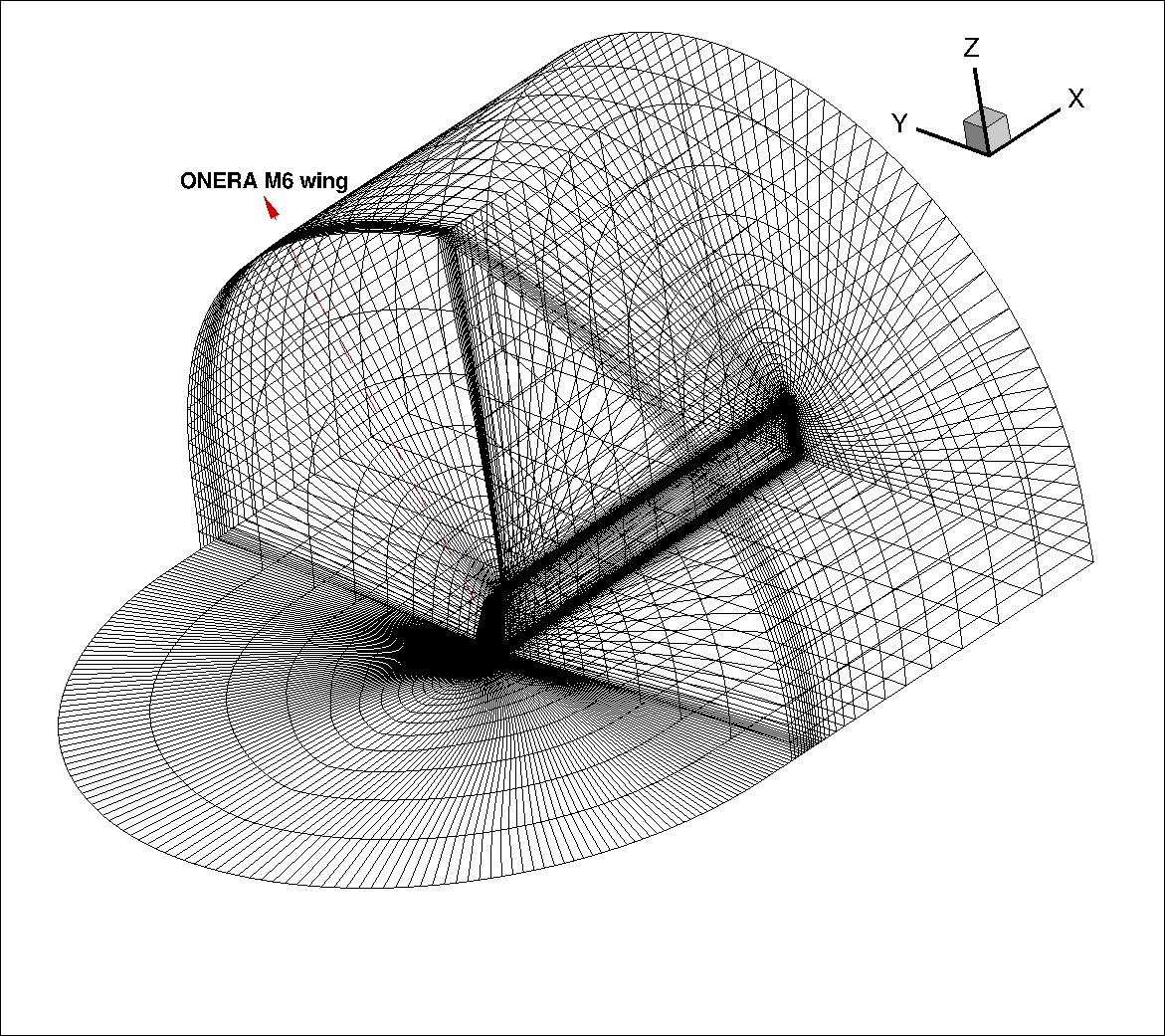} 
	} 
	\subfigure[The domain decomposition for the ONERA M6 wing case using 16 GPUs]{ 
		\label{m6onera-id}
		\includegraphics[width=.45\textwidth,trim=7 7 7 7,clip]{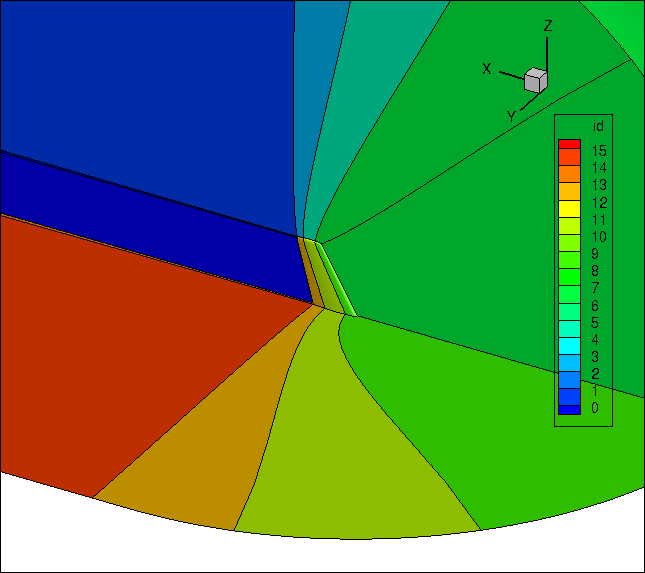} 
	} 
	\caption{Grid and domain decomposition for ONERA M6 wing} 
	\label{m6onera_grid_id}
\end{figure}

\begin{figure}[H]
	\centering 
	\subfigure[The relative residual norm history for ONERA M6 wing]{ 
		\label{m6onera-resid}
		\includegraphics[width=.45\textwidth,trim=7 7 7 7,clip]{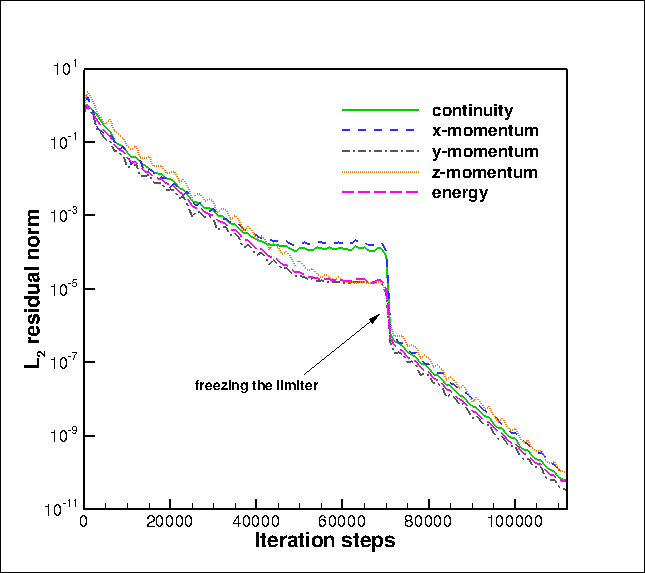} 
	} 
	\subfigure[The laminar NS pressure coefficient contour for ONERA M6 wing]{ 
		\label{m6onera_cp_NS_2}
		\includegraphics[width=.45\textwidth,trim=7 7 7 7,clip]{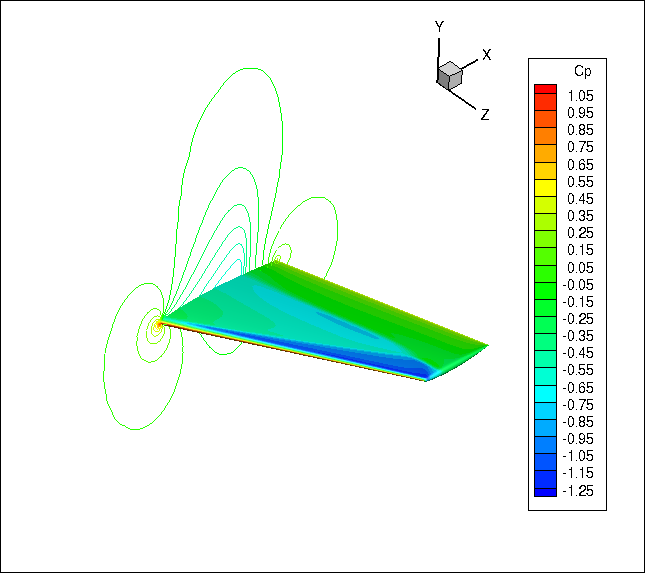} 
	} 
	\caption{Residual history and solution for ONERA M6 wing} 
	\label{m6onera_residual_solution}
\end{figure}

The parallel solution and the serial solution of the wing case have been compared to each other on a coarse mesh and the relative errors for primitive variables based on the farfield boundary values is within round-off error ($10^{-12}$).A coarse level of grid and the domain decomposition of using 16 GPUs are given in Fig.~\ref{m6onera_grid} and Fig.~\ref{m6onera-id}, respectively. The relative iterative residual $L_2$ norm history and the pressure coefficient ($C_p$) contour using the laminar NS solver in SENSEI are given in Fig.~\ref{m6onera-resid} and Fig.\ref{m6onera_cp_NS_2}, respectively. From Fig.~\ref{m6onera-resid}, it can be seen that the iterative errors have been driven down small enough. 

\begin{figure}[H]
	\centering
	\includegraphics[width=.7\textwidth,trim=7 7 7 7,clip]{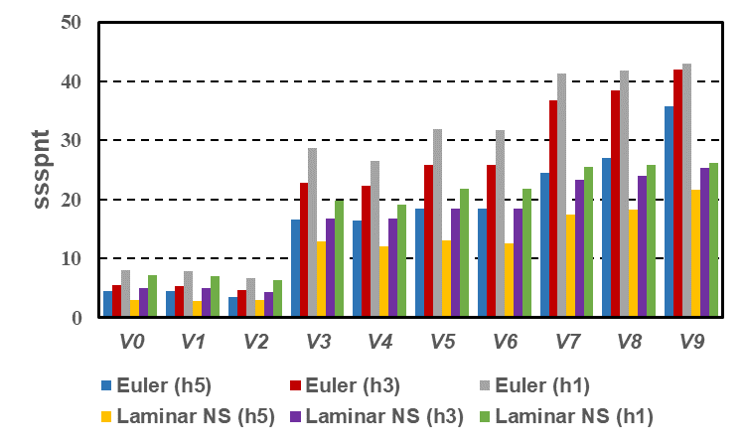}
	\caption{The single P100 GPU performance of different versions for ONERA M6 wing}
	\label{onera-versions}
\end{figure}

Since this wing case is 3D and has multiple parent blocks, we are interested in whether the performance optimizations introduced earlier can improve the performance of this wing case. Fig.~\ref{onera-versions} shows the performance of different versions for the ONERA M6 wing case. From the grid level of $h5$ to $h1$, the grid refinement factor is 2 (refined in $z$, $y$ and $x$ cyclically). $V2$ runs slower than $V1$ for all levels of grid, indicating that the extrapolation to ghost cells on the GPU is not as efficient as that on the CPU, although it is parallelized. With proper optimization, $V3$ is about 3 to 4 times faster than $V2$, which is similar to the previous two 2D cases. From $V3$ to $V4$, there is a performance drop for almost all runs, no matter what the grid level and the solver is. Splitting one kernel into two kernels for this case incurs some overhead and reduces the compute utilization a bit. There is a slight performance improvement from $V4$ to $V5$ when using the derived type to buffer the boundary data for connected boundaries. The data will be allocated in the main memory of the GPU before needed, which outperforms the use of dynamic data to buffer the boundary data. For a single GPU, $V5$ and $V6$ perform equivalently fast. Further performance optimization on the boundary flux calculation can improve the performance significantly, which can be seen from $V6$ to $V7$. Carefully moving the data between the host and the device can improve the performance on coarse levels of grid, but not on very fine levels of grid, as the computation becomes more dominant when refining the grid.

\begin{figure}[H]
	\centering
	\includegraphics[width=.7\textwidth,trim=7 7 7 7,clip]{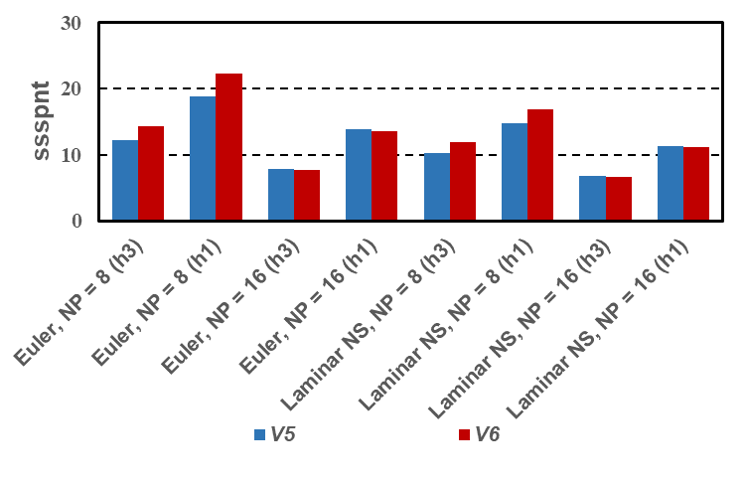}
	\caption{Performance comparison between $V5$ and $V6$ for ONERA M6 wing (P100 GPU)}
	\label{onera-V5V6}
\end{figure}

Although the wing case has multiple parent blocks, there is no MPI communication if using only a single GPU. Therefore, there is only negligible difference between $V5$ and $V6$. Similar to the NACA 0012 airfoil case, multiple GPUs are used to show the effect of reordering the non-blocking MPI I\_send/I\_recv calls and the MPI\_Wait calls. Fig.~\ref{onera-V5V6} shows that there are some performance gains for some runs but not all. $V6$ accelerates the code by 14\% to 18\% when $np$ is equal to 8. If using 16 GPUs, more connected boundaries are created, and it impedes the performance improvement. A possible reason for this may be that although the implementation from $V5$ to $V6$ exposes more asynchronous progression on the implementation side, the platform communication system does not support that very well when too many communication calls are invoked. This issue may be resolved if switching to the MPI+OpenACC+OpenMP model, which is not covered in this paper. However, it can be seen that the performance degradation using 16 GPUs is only 0.8\% to 3\%, which is small.

\begin{figure}[H]
	\centering 
	\subfigure[Strong scaling (Euler)]{ 
		\label{euler-onera-strong}
		\includegraphics[width=.45\textwidth,trim=7 7 0 7,clip]{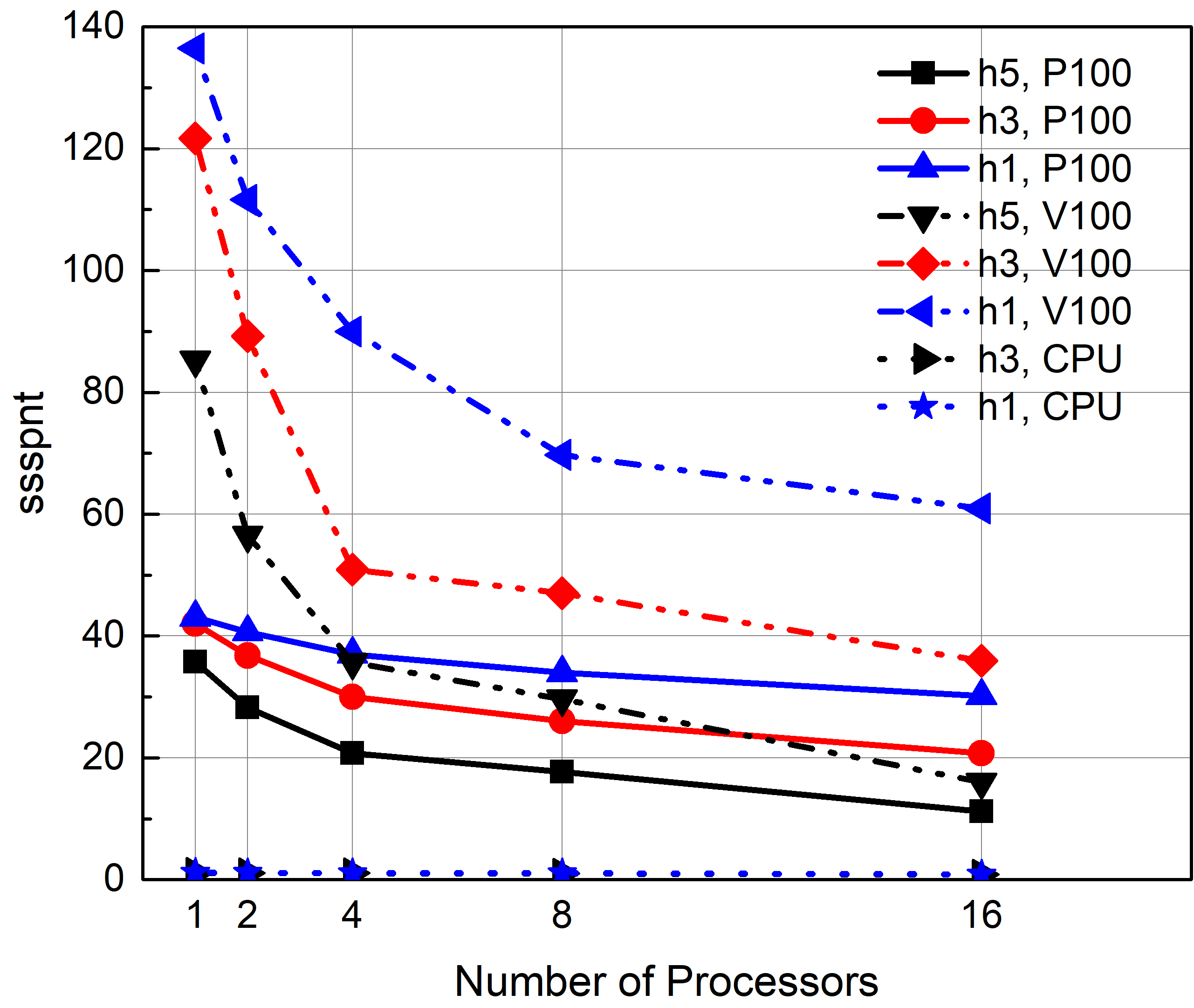} 
	} 
	\subfigure[Weak scaling (Euler)]{ 
		\label{euler-onera-weak}
		\includegraphics[width=.45\textwidth,trim=7 7 0 7,clip]{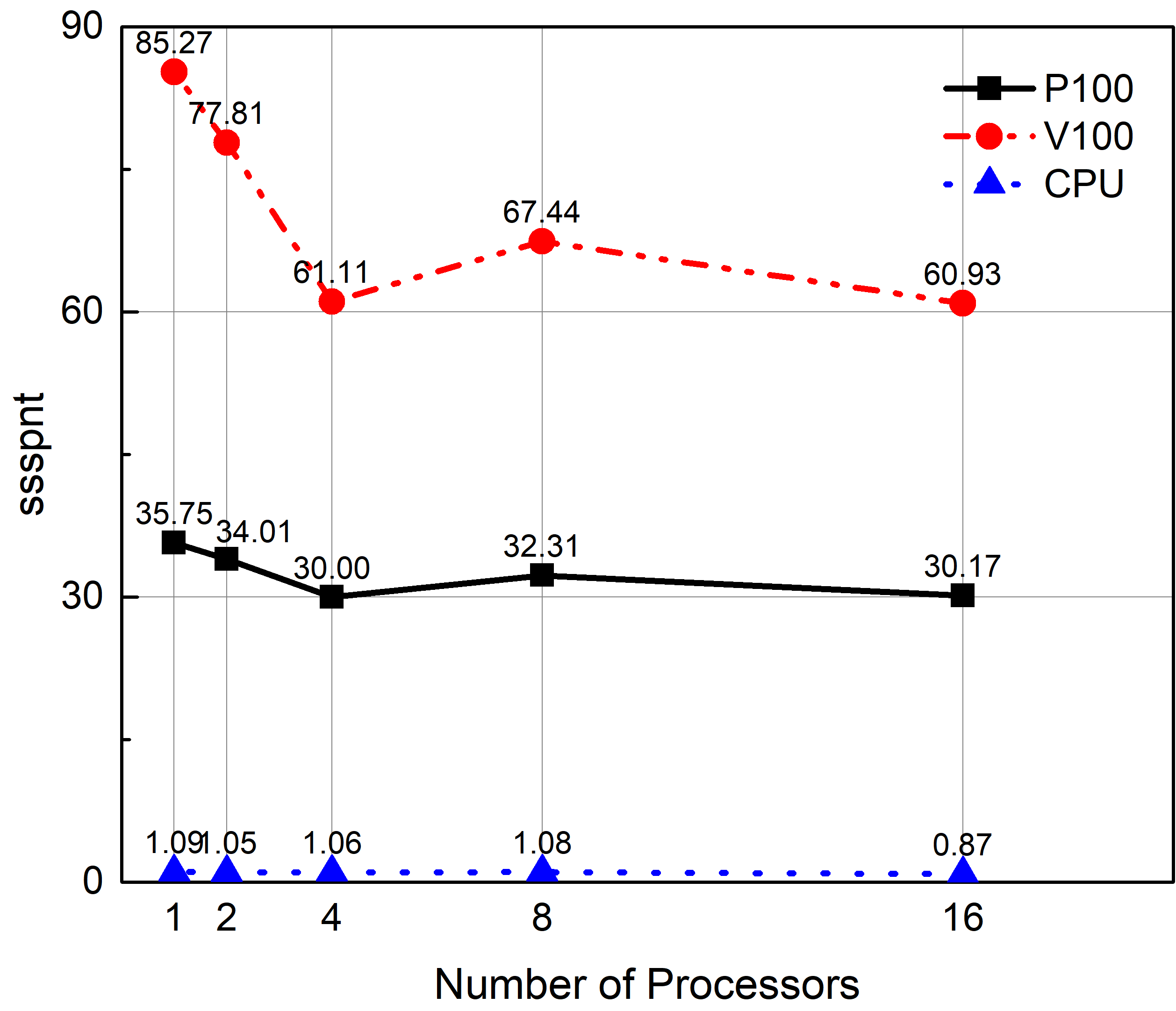} 
	} 
	\caption{The scaling performance for the 3D Euler ONERA M6 wing case} 
	\label{Euler_ONERA}
\end{figure}

\begin{figure}[H]
	\centering 
	\subfigure[Strong scaling (laminar NS)]{ 
		\label{NS-onera-strong}
		\includegraphics[width=.45\textwidth,trim=7 7 0 7,clip]{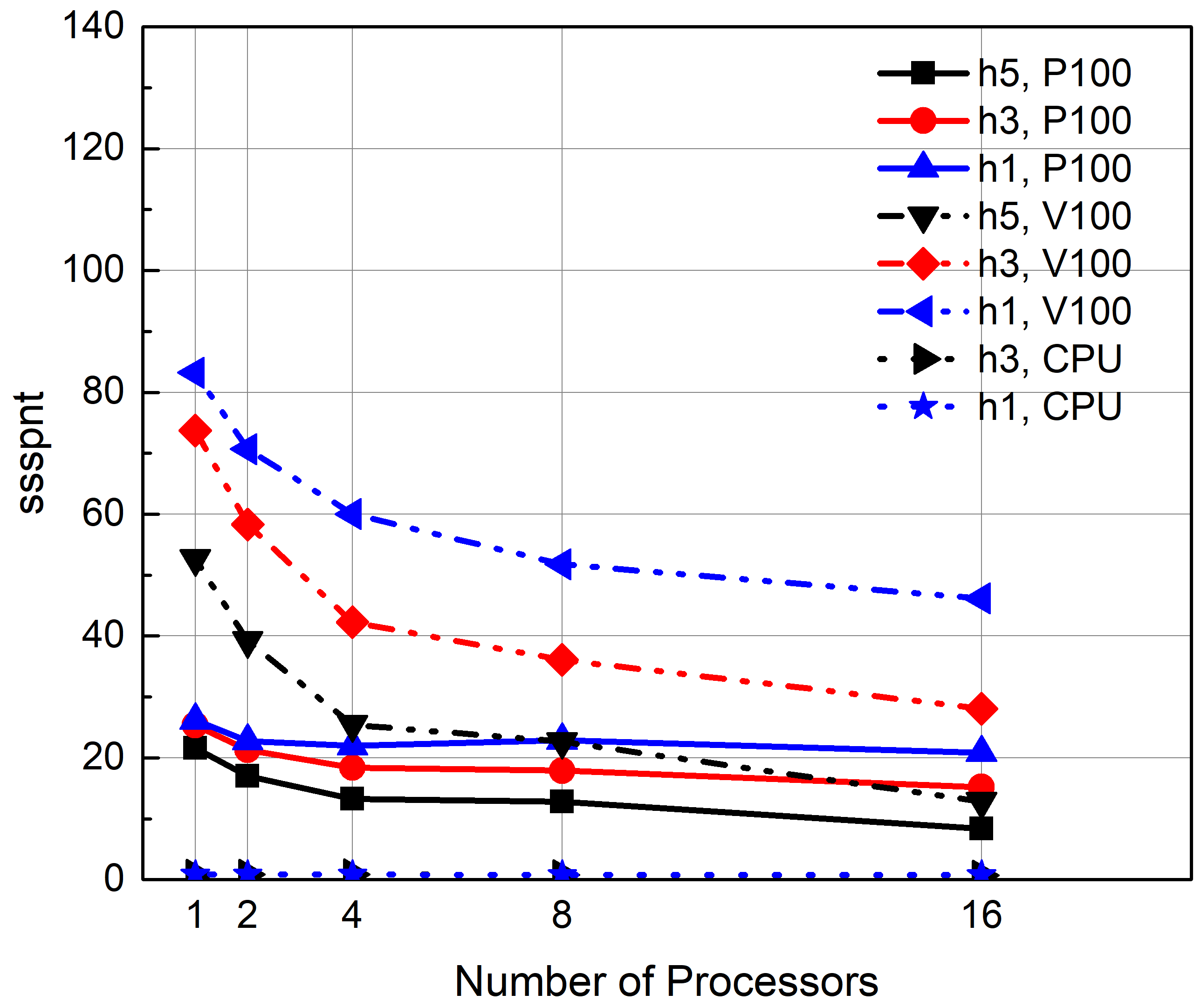} 
	} 
	\subfigure[Weak scaling (laminar NS)]{ 
		\label{NS-onera-weak}
		\includegraphics[width=.45\textwidth,trim=7 7 0 7,clip]{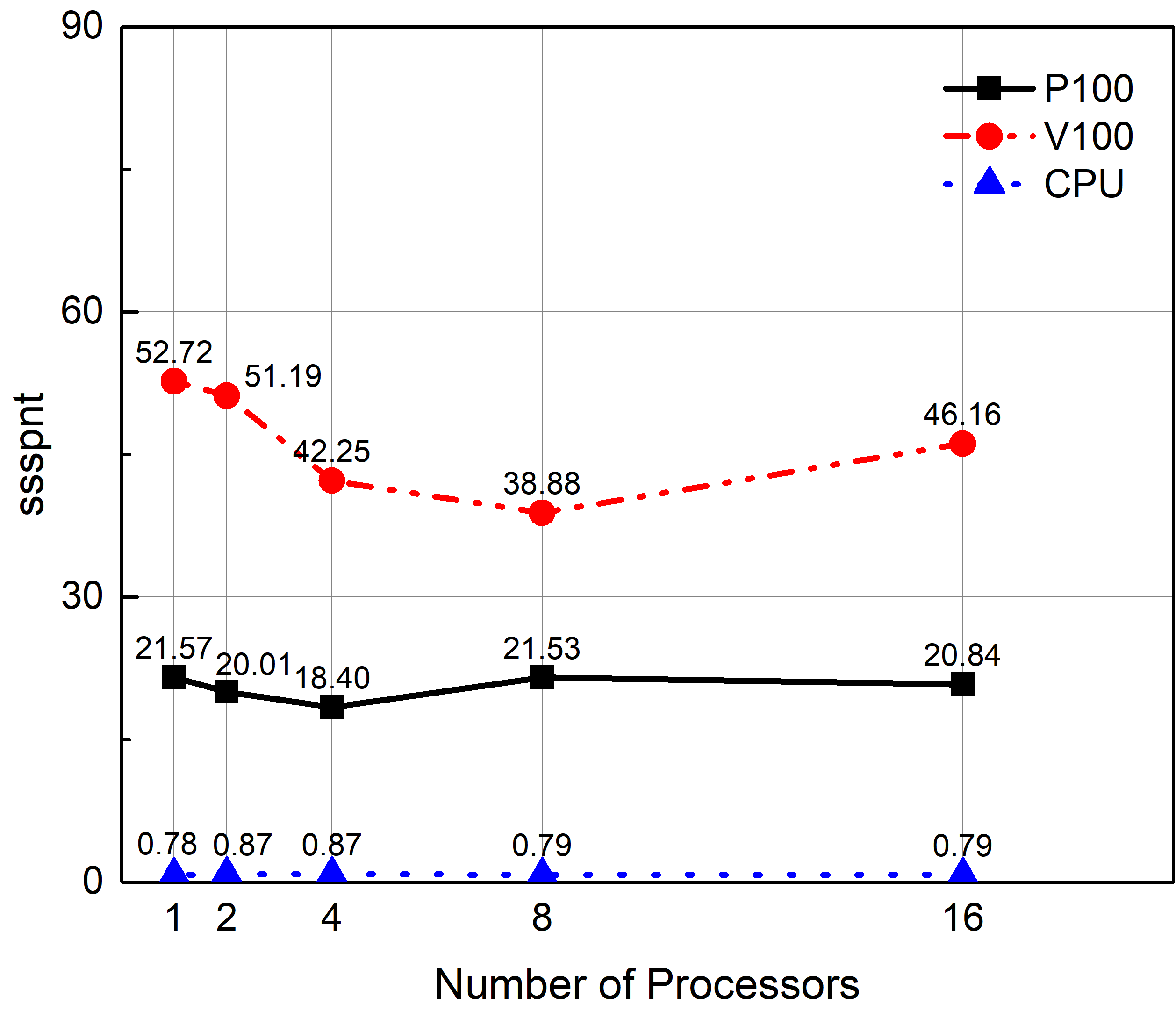} 
	} 
	\caption{The scaling performance for the 3D laminar NS ONERA M6 wing case} 
	\label{NS_ONERA}
\end{figure}

Fig.~\ref{Euler_ONERA} and Fig.~\ref{NS_ONERA} show the strong and weak scaling performance using Euler and laminar NS solvers, respectively. Some different behaviours show as in this case some processors need to hold multiple blocks, which is different from the 2D inlet and 2D NACA 0012 case. A single GPU is about 33 times faster than a single CPU on the $h5$ level grid. The weak scaling of the GPU keeps good efficiency over the whole $np$ range shown in Fig.~\ref{euler-onera-weak} and Fig.~\ref{NS-onera-weak}.

\subsection{GPUDirect}

\begin{figure}[H]
	\centering 
	\subfigure[Subsonic flow past a NACA 0012 airfoil]{ 
		\label{naca0012-V9V10}
		\includegraphics[width=.45\textwidth,trim=7 7 0 7,clip]{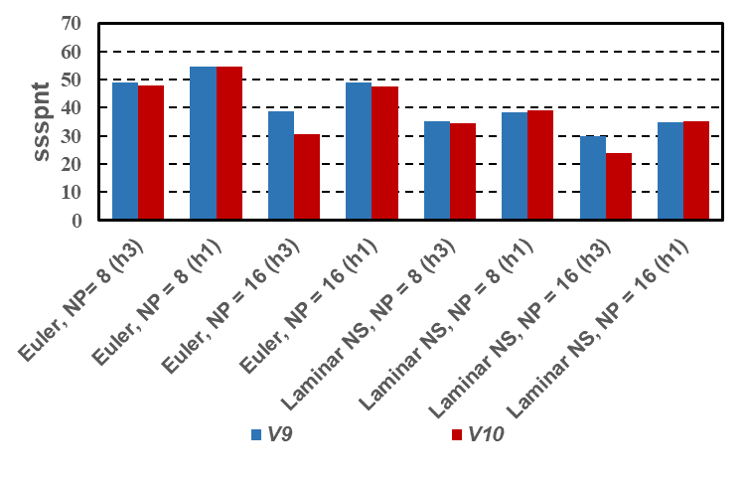} 
	} 
	\subfigure[Transonic flow past an ONERA M6 wing]{ 
		\label{onera-V9V10}
		\includegraphics[width=.45\textwidth,trim=7 7 0 7,clip]{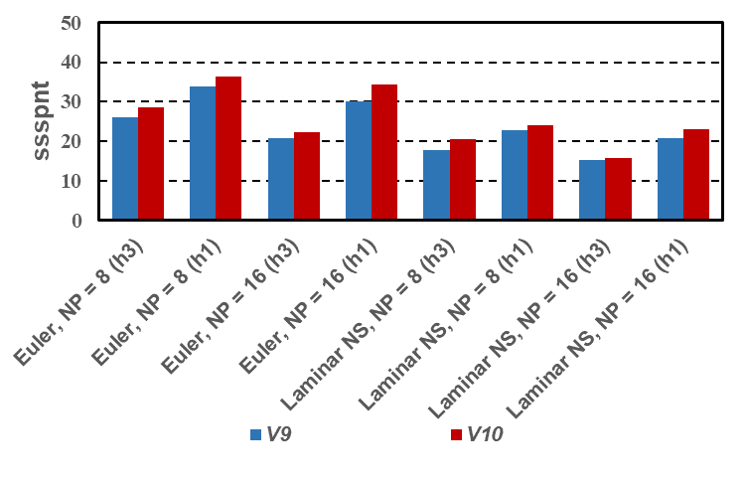} 
	} 
	\caption{Performance comparison between $V9$ and $V10$} 
	\label{V9-V10}
\end{figure}

Since GPUDirect is not a general performance optimization, as it requires some support from both the compiler side and the communication system side, a comparison of $V9$ and $V10$ is made at the end to give readers more insights of the effect of GPUDirect. GPUDirect was applied to the 2D Euler/laminar flow past the NACA 0012 airfoil and the transonic flow over the 3D ONERA M6 wing. It should be noted that there is no guarantee that using GPUDirect can improve the performance substantially without the hardware support like using NVLink (however both the NewRiver and the Cascades cluster does not have NVLink so the memory bandwidth is still not high enough). It can be found that the two cases show different behaviours when applying GPUDirect, seen in Fig.~\ref{V9-V10}. For the NACA 0012 case, generally $V10$ is slower than $V9$, which means that GPUDirect makes the code to run slower. However, for the ONERA wing case, using GPUDirect improves the performance by 4\% to 14\%. Whether there is a performance gain or not depends on the problem and number of communications. Commonly if high memory bandwidth NVLink is available, GPUDirect should be more beneficial to the performance.

\section{Conclusions \& Future Work}

An improved framework using MPI+OpenACC is developed to accelerate a CFD code on multi-block structured grids. OpenACC has some advantages in terms of the ease of programming, the good portability and the fair performance. A processor-clustered domain decomposition and a block-clustered domain aggregation method are used to balance the workload among processors. Also, the communication overhead is not high using the domain decomposition and aggregation methods. A parallel boundary decomposition method is also proposed with the use of the MPI inter-communicator functions. The boundary reordering for multi-block cases is addressed to avoid the dead lock issue when sending and receiving messages. A number of performance optimizations are examined, such as using the global derived type to buffer the connected boundary data, removing temporary arrays when making procedure calls, reordering of blocking calls for non-blocking MPI communications for multi-block cases, using GPUDirect, etc. These performance optimizations have been demonstrated to improve single GPU performance more than up to 5 times compared to the baseline GPU version. More importantly, all the three test cases show good strong and weak scaling up to 16 GPUs, with a good parallel efficiency if the problem is large enough.

\bibliography{mybib}

\begin{thebibliography}{58}
\expandafter\ifx\csname natexlab\endcsname\relax\def\natexlab#1{#1}\fi
\providecommand{\url}[1]{\texttt{#1}}
\providecommand{\href}[2]{#2}
\providecommand{\path}[1]{#1}
\providecommand{\DOIprefix}{doi:}
\providecommand{\ArXivprefix}{arXiv:}
\providecommand{\URLprefix}{URL: }
\providecommand{\Pubmedprefix}{pmid:}
\providecommand{\doi}[1]{\href{http://dx.doi.org/#1}{\path{#1}}}
\providecommand{\Pubmed}[1]{\href{pmid:#1}{\path{#1}}}
\providecommand{\bibinfo}[2]{#2}
\ifx\xfnm\relax \def\xfnm[#1]{\unskip,\space#1}\fi
\bibitem[{Barney(2020)}]{parallelcomputing}
\bibinfo{author}{B.~Barney}, \bibinfo{title}{{Introduction to Parallel
  Computing}}, \bibinfo{year}{2020}. \URLprefix
  \url{https://computing.llnl.gov/tutorials/parallel_comp/},
  \bibinfo{note}{(last accessed on 07/24/20)}.
\bibitem[{Landaverde et~al.(2014)Landaverde, Zhang, Coskun, and
  Herbordt}]{landaverde2014investigation}
\bibinfo{author}{R.~Landaverde}, \bibinfo{author}{T.~Zhang},
  \bibinfo{author}{A.~K. Coskun}, \bibinfo{author}{M.~Herbordt},
\newblock \bibinfo{title}{{An Investigation of Unified Memory Access
  Performance in CUDA}},
\newblock in: \bibinfo{booktitle}{2014 IEEE High Performance Extreme Computing
  Conference (HPEC)}, \bibinfo{organization}{IEEE}, \bibinfo{address}{Waltham,
  MA, US}, \bibinfo{year}{2014}, pp. \bibinfo{pages}{1--6}.
\bibitem[{Barney(2020{\natexlab{a}})}]{OpenMP2}
\bibinfo{author}{B.~Barney}, \bibinfo{title}{{OpenMP}},
  \bibinfo{year}{2020}{\natexlab{a}}. \URLprefix
  \url{https://computing.llnl.gov/tutorials/openMP/}, \bibinfo{note}{(last
  accessed on 07/24/20)}.
\bibitem[{Barney(2020{\natexlab{b}})}]{MPI2}
\bibinfo{author}{B.~Barney}, \bibinfo{title}{{Message Passing Interface
  (MPI)}}, \bibinfo{year}{2020}{\natexlab{b}}. \URLprefix
  \url{https://computing.llnl.gov/tutorials/mpi/}, \bibinfo{note}{(last
  accessed on 07/24/20)}.
\bibitem[{NVIDIA(2019)}]{CUDA}
\bibinfo{author}{NVIDIA}, \bibinfo{title}{{CUDA C++ Programming Guide}},
  \bibinfo{year}{2019}. \URLprefix
  \url{https://docs.nvidia.com/pdf/CUDA_C_Programming_Guide.pdf},
  \bibinfo{note}{(last accessed on 07/24/20)}.
\bibitem[{{Khronos OpenCL Working Group}(2019)}]{OpenCL}
\bibinfo{author}{{Khronos OpenCL Working Group}}, \bibinfo{title}{{The OpenCL C
  2.0 Specification}}, \bibinfo{year}{2019}. \URLprefix
  \url{https://www.khronos.org/registry/OpenCL/specs/2.2/pdf/OpenCL_C.pdf},
  \bibinfo{note}{(last accessed on 07/24/20)}.
\bibitem[{Ope(2015)}]{OpenACC}
\bibinfo{title}{{OpenACC Programming and Best Practices Guide}},
  \bibinfo{year}{2015}. \URLprefix
  \url{https://www.openacc.org/sites/default/files/inline-files/OpenACC_Programming_Guide_0.pdf},
  \bibinfo{note}{(last accessed on 07/24/20)}.
\bibitem[{Gourdain et~al.(2009{\natexlab{a}})Gourdain, Gicquel, Montagnac,
  Vermorel, Gazaix, Staffelbach, Garcia, Boussuge, and
  Poinsot}]{gourdain2009high}
\bibinfo{author}{N.~Gourdain}, \bibinfo{author}{L.~Gicquel},
  \bibinfo{author}{M.~Montagnac}, \bibinfo{author}{O.~Vermorel},
  \bibinfo{author}{M.~Gazaix}, \bibinfo{author}{G.~Staffelbach},
  \bibinfo{author}{M.~Garcia}, \bibinfo{author}{J.~Boussuge},
  \bibinfo{author}{T.~Poinsot},
\newblock \bibinfo{title}{{High Performance Parallel Computing of Flows in
  Complex Geometries: I. Methods}},
\newblock \bibinfo{journal}{Computational Science \& Discovery}
  \bibinfo{volume}{2} (\bibinfo{year}{2009}{\natexlab{a}})
  \bibinfo{pages}{015003}.
\bibitem[{Gourdain et~al.(2009{\natexlab{b}})Gourdain, Gicquel, Staffelbach,
  Vermorel, Duchaine, Boussuge, and Poinsot}]{gourdain2009high2}
\bibinfo{author}{N.~Gourdain}, \bibinfo{author}{L.~Gicquel},
  \bibinfo{author}{G.~Staffelbach}, \bibinfo{author}{O.~Vermorel},
  \bibinfo{author}{F.~Duchaine}, \bibinfo{author}{J.~Boussuge},
  \bibinfo{author}{T.~Poinsot},
\newblock \bibinfo{title}{{High Performance Parallel Computing of Flows in
  Complex Geometries: II. Applications}},
\newblock \bibinfo{journal}{Computational Science \& Discovery}
  \bibinfo{volume}{2} (\bibinfo{year}{2009}{\natexlab{b}})
  \bibinfo{pages}{015004}.
\bibitem[{Amritkar et~al.(2014)Amritkar, Deb, and
  Tafti}]{amritkar2014efficient}
\bibinfo{author}{A.~Amritkar}, \bibinfo{author}{S.~Deb},
  \bibinfo{author}{D.~Tafti},
\newblock \bibinfo{title}{{Efficient Parallel CFD-DEM Simulations using
  OpenMP}},
\newblock \bibinfo{journal}{Journal of Computational Physics}
  \bibinfo{volume}{256} (\bibinfo{year}{2014}) \bibinfo{pages}{501--519}.
\bibitem[{Krpic et~al.(2012)Krpic, Martinovic, and Crnkovic}]{krpic2012green}
\bibinfo{author}{Z.~Krpic}, \bibinfo{author}{G.~Martinovic},
  \bibinfo{author}{I.~Crnkovic},
\newblock \bibinfo{title}{{Green HPC: MPI vs. OpenMP on a Shared Memory
  System}},
\newblock in: \bibinfo{booktitle}{2012 Proceedings of the 35th International
  Convention MIPRO}, \bibinfo{organization}{IEEE}, \bibinfo{address}{Opatija,
  Croatia}, \bibinfo{year}{2012}, pp. \bibinfo{pages}{246--250}.
\bibitem[{Mininni et~al.(2011)Mininni, Rosenberg, Reddy, and
  Pouquet}]{mininni2011hybrid}
\bibinfo{author}{P.~D. Mininni}, \bibinfo{author}{D.~Rosenberg},
  \bibinfo{author}{R.~Reddy}, \bibinfo{author}{A.~Pouquet},
\newblock \bibinfo{title}{{A Hybrid MPI--OpenMP Scheme for Scalable Parallel
  Pseudospectral Computations for Fluid Turbulence}},
\newblock \bibinfo{journal}{Parallel Computing} \bibinfo{volume}{37}
  (\bibinfo{year}{2011}) \bibinfo{pages}{316--326}.
\bibitem[{Owens et~al.(2008)Owens, Houston, Luebke, Green, Stone, and
  Phillips}]{owens2008gpu}
\bibinfo{author}{J.~D. Owens}, \bibinfo{author}{M.~Houston},
  \bibinfo{author}{D.~Luebke}, \bibinfo{author}{S.~Green},
  \bibinfo{author}{J.~E. Stone}, \bibinfo{author}{J.~C. Phillips},
\newblock \bibinfo{title}{{GPU Computing}},
\newblock \bibinfo{journal}{Proceedings of the IEEE} \bibinfo{volume}{96}
  (\bibinfo{year}{2008}) \bibinfo{pages}{879--899}.
\bibitem[{Herdman et~al.(2012)Herdman, Gaudin, McIntosh-Smith, Boulton,
  Beckingsale, Mallinson, and Jarvis}]{herdman2012accelerating}
\bibinfo{author}{J.~Herdman}, \bibinfo{author}{W.~Gaudin},
  \bibinfo{author}{S.~McIntosh-Smith}, \bibinfo{author}{M.~Boulton},
  \bibinfo{author}{D.~A. Beckingsale}, \bibinfo{author}{A.~Mallinson},
  \bibinfo{author}{S.~A. Jarvis},
\newblock \bibinfo{title}{{Accelerating Hydrocodes with OpenACC, OpenCL and
  CUDA}},
\newblock in: \bibinfo{booktitle}{2012 SC Companion: High Performance
  Computing, Networking Storage and Analysis}, \bibinfo{organization}{IEEE},
  \bibinfo{address}{Salt Lake City, UT, USA}, \bibinfo{year}{2012}, pp.
  \bibinfo{pages}{465--471}.
\bibitem[{Jacobsen and Senocak(2013)}]{jacobsen2013multi}
\bibinfo{author}{D.~A. Jacobsen}, \bibinfo{author}{I.~Senocak},
\newblock \bibinfo{title}{{Multi-level Parallelism for Incompressible Flow
  Computations on GPU Clusters}},
\newblock \bibinfo{journal}{Parallel Computing} \bibinfo{volume}{39}
  (\bibinfo{year}{2013}) \bibinfo{pages}{1--20}.
\bibitem[{Elsen et~al.(2008)Elsen, LeGresley, and Darve}]{elsen2008large}
\bibinfo{author}{E.~Elsen}, \bibinfo{author}{P.~LeGresley},
  \bibinfo{author}{E.~Darve},
\newblock \bibinfo{title}{{Large Calculation of the Flow over a Hypersonic
  Vehicle using a GPU}},
\newblock \bibinfo{journal}{Journal of Computational Physics}
  \bibinfo{volume}{227} (\bibinfo{year}{2008}) \bibinfo{pages}{10148--10161}.
\bibitem[{Buck et~al.(2004)Buck, Foley, Horn, Sugerman, Fatahalian, Houston,
  and Hanrahan}]{buck2004brook}
\bibinfo{author}{I.~Buck}, \bibinfo{author}{T.~Foley},
  \bibinfo{author}{D.~Horn}, \bibinfo{author}{J.~Sugerman},
  \bibinfo{author}{K.~Fatahalian}, \bibinfo{author}{M.~Houston},
  \bibinfo{author}{P.~Hanrahan},
\newblock \bibinfo{title}{{Brook for GPUs: Stream Computing on Graphics
  Hardware}},
\newblock \bibinfo{journal}{ACM Transactions on Graphics (TOG)}
  \bibinfo{volume}{23} (\bibinfo{year}{2004}) \bibinfo{pages}{777--786}.
\bibitem[{Brandvik and Pullan(2008)}]{brandvik2008acceleration}
\bibinfo{author}{T.~Brandvik}, \bibinfo{author}{G.~Pullan},
\newblock \bibinfo{title}{{Acceleration of a 3D Euler Solver using Commodity
  Graphics Hardware}},
\newblock in: \bibinfo{booktitle}{46th AIAA Aerospace Sciences Meeting and
  Exhibit}, \bibinfo{address}{Reno, Nevada, US}, \bibinfo{year}{2008}, p.
  \bibinfo{pages}{607}.
\bibitem[{Luo et~al.(2013)Luo, Edwards, Luo, and Mueller}]{luo2013performance}
\bibinfo{author}{L.~Luo}, \bibinfo{author}{J.~R. Edwards},
  \bibinfo{author}{H.~Luo}, \bibinfo{author}{F.~Mueller},
\newblock \bibinfo{title}{{Performance Assessment of a Multiblock
  Incompressible Navier-Stokes Solver using Directive-based GPU Programming in
  a Cluster Environment}},
\newblock in: \bibinfo{booktitle}{52nd Aerospace Sciences Meeting},
  \bibinfo{address}{National Harbor, MD, US}, \bibinfo{year}{2013}.
\bibitem[{Xia et~al.(2015)Xia, Lou, Luo, Edwards, and Mueller}]{xia2015openacc}
\bibinfo{author}{Y.~Xia}, \bibinfo{author}{J.~Lou}, \bibinfo{author}{H.~Luo},
  \bibinfo{author}{J.~Edwards}, \bibinfo{author}{F.~Mueller},
\newblock \bibinfo{title}{{OpenACC Acceleration of an Unstructured CFD Solver
  based on a Reconstructed Discontinuous Galerkin Method for Compressible
  Flows}},
\newblock \bibinfo{journal}{International Journal for Numerical Methods in
  Fluids} \bibinfo{volume}{78} (\bibinfo{year}{2015})
  \bibinfo{pages}{123--139}.
\bibitem[{Chandar et~al.(2013)Chandar, Sitaraman, and
  Mavriplis}]{chandar2013hybrid}
\bibinfo{author}{D.~D. Chandar}, \bibinfo{author}{J.~Sitaraman},
  \bibinfo{author}{D.~J. Mavriplis},
\newblock \bibinfo{title}{{A Hybrid Multi-GPU/CPU Computational Framework for
  Rotorcraft Flows on Unstructured Overset Grids}},
\newblock in: \bibinfo{booktitle}{21st AIAA Computational Fluid Dynamics
  Conference}, \bibinfo{address}{San Diego, CA, US}, \bibinfo{year}{2013}, p.
  \bibinfo{pages}{2855}.
\bibitem[{Xue et~al.(2018)Xue, Jackson, and Roy}]{xue2018multi}
\bibinfo{author}{W.~Xue}, \bibinfo{author}{C.~W. Jackson},
  \bibinfo{author}{C.~J. Roy},
\newblock \bibinfo{title}{{Multi-CPU/GPU Parallelization, Optimization and
  Machine Learning based Autotuning of Structured Grid CFD Codes}},
\newblock in: \bibinfo{booktitle}{2018 AIAA Aerospace Sciences Meeting},
  \bibinfo{address}{Kissimmee, FL, US}, \bibinfo{year}{2018}, p.
  \bibinfo{pages}{0362}.
\bibitem[{Xue and Roy(2020{\natexlab{a}})}]{xue2020multi}
\bibinfo{author}{W.~Xue}, \bibinfo{author}{C.~J. Roy},
\newblock \bibinfo{title}{{Multi-GPU Performance Optimization of a
  Computational Fluid Dynamics Code using OpenACC}},
\newblock \bibinfo{journal}{Concurrency and Computation: Practice and
  Experience}  (\bibinfo{year}{2020}{\natexlab{a}}) \bibinfo{pages}{e6036}.
\bibitem[{Xue and Roy(2020{\natexlab{b}})}]{xue2020heterogeneous}
\bibinfo{author}{W.~Xue}, \bibinfo{author}{C.~J. Roy},
\newblock \bibinfo{title}{{Heterogeneous Computing of CFD Applications on
  CPU-GPU Platforms using OpenACC Directives}},
\newblock in: \bibinfo{booktitle}{AIAA Scitech 2020 Forum},
  \bibinfo{address}{Orlando, FL, US}, \bibinfo{year}{2020}{\natexlab{b}}, p.
  \bibinfo{pages}{1046}.
\bibitem[{Derlaga et~al.(2013)Derlaga, Phillips, and Roy}]{derlaga2013sensei}
\bibinfo{author}{J.~M. Derlaga}, \bibinfo{author}{T.~Phillips},
  \bibinfo{author}{C.~J. Roy},
\newblock \bibinfo{title}{{SENSEI Computational Fluid Dynamics Code: A Case
  Study in Modern Fortran Software Development}},
\newblock in: \bibinfo{booktitle}{21st AIAA Computational Fluid Dynamics
  Conference}, \bibinfo{address}{San Diego, CA, US}, \bibinfo{year}{2013}.
\bibitem[{Jackson et~al.(2019)Jackson, Tyson, and Roy}]{jackson2019turbulence}
\bibinfo{author}{C.~W. Jackson}, \bibinfo{author}{W.~C. Tyson},
  \bibinfo{author}{C.~J. Roy},
\newblock \bibinfo{title}{{Turbulence Model Implementation and Verification in
  the SENSEI CFD Code}},
\newblock in: \bibinfo{booktitle}{AIAA Scitech 2019 Forum},
  \bibinfo{address}{San Diego, CA}, \bibinfo{year}{2019}, p.
  \bibinfo{pages}{2331}.
\bibitem[{Xue et~al.(2020)Xue, Wang, and Roy}]{xue2020code}
\bibinfo{author}{W.~Xue}, \bibinfo{author}{H.~Wang}, \bibinfo{author}{C.~J.
  Roy},
\newblock \bibinfo{title}{{Code Verification for 3D Turbulence Modeling in
  Parallel SENSEI Accelerated with MPI}},
\newblock in: \bibinfo{booktitle}{AIAA Scitech 2020 Forum},
  \bibinfo{address}{Orlando, FL, US}, \bibinfo{year}{2020}, p.
  \bibinfo{pages}{0347}.
\bibitem[{Oberkampf and Roy(2010)}]{oberkampf2010verification}
\bibinfo{author}{W.~L. Oberkampf}, \bibinfo{author}{C.~J. Roy},
  \bibinfo{title}{Verification and validation in scientific computing},
  \bibinfo{publisher}{Cambridge University Press}, \bibinfo{year}{2010}.
\bibitem[{Van~Leer(1979)}]{van1979towards}
\bibinfo{author}{B.~Van~Leer},
\newblock \bibinfo{title}{Towards the ultimate conservative difference scheme.
  v. a second-order sequel to godunov's method},
\newblock \bibinfo{journal}{Journal of computational Physics}
  \bibinfo{volume}{32} (\bibinfo{year}{1979}) \bibinfo{pages}{101--136}.
\bibitem[{Jameson et~al.(1981)Jameson, Schmidt, and
  Turkel}]{jameson1981numerical}
\bibinfo{author}{A.~Jameson}, \bibinfo{author}{W.~Schmidt},
  \bibinfo{author}{E.~Turkel},
\newblock \bibinfo{title}{{Numerical Solution of the Euler Equations by Finite
  Volume Methods using Runge Kutta Time Stepping Schemes}},
\newblock in: \bibinfo{booktitle}{14th fluid and plasma dynamics conference},
  \bibinfo{address}{Palo Alto, CA, US}, \bibinfo{year}{1981}, p.
  \bibinfo{pages}{1259}.
\bibitem[{Ascher et~al.(1997)Ascher, Ruuth, and Spiteri}]{ascher1997implicit}
\bibinfo{author}{U.~M. Ascher}, \bibinfo{author}{S.~J. Ruuth},
  \bibinfo{author}{R.~J. Spiteri},
\newblock \bibinfo{title}{Implicit-explicit runge-kutta methods for
  time-dependent partial differential equations},
\newblock \bibinfo{journal}{Applied Numerical Mathematics} \bibinfo{volume}{25}
  (\bibinfo{year}{1997}) \bibinfo{pages}{151--167}.
\bibitem[{Kennedy and Carpenter(2016)}]{kennedy2016diagonally}
\bibinfo{author}{C.~A. Kennedy}, \bibinfo{author}{M.~H. Carpenter},
\newblock \bibinfo{title}{{Diagonally Implicit Runge-Kutta Methods for Ordinary
  Differential Equations. A Review}}  (\bibinfo{year}{2016}).
\bibitem[{Wu et~al.(1990)Wu, Fan, and Erickson}]{wu1990three}
\bibinfo{author}{J.~Wu}, \bibinfo{author}{L.~Fan},
  \bibinfo{author}{L.~Erickson},
\newblock \bibinfo{title}{{Three-Point Backward Finite-Difference Method for
  Solving a System of Mixed Hyperbolic—Parabolic Partial Differential
  Equations}},
\newblock \bibinfo{journal}{Computers \& Chemical Engineering}
  \bibinfo{volume}{14} (\bibinfo{year}{1990}) \bibinfo{pages}{679--685}.
\bibitem[{Roe(1981)}]{roe1981approximate}
\bibinfo{author}{P.~L. Roe},
\newblock \bibinfo{title}{{Approximate Riemann Solvers, Parameter Vectors, and
  Difference Schemes}},
\newblock \bibinfo{journal}{Journal of Computational Physics}
  \bibinfo{volume}{43} (\bibinfo{year}{1981}) \bibinfo{pages}{357--372}.
\bibitem[{Steger and Warming(1981)}]{steger1981flux}
\bibinfo{author}{J.~L. Steger}, \bibinfo{author}{R.~Warming},
\newblock \bibinfo{title}{{Flux Vector Splitting of the Inviscid Gasdynamic
  Equations with Application to Finite-Difference Methods}},
\newblock \bibinfo{journal}{Journal of Computational Physics}
  \bibinfo{volume}{40} (\bibinfo{year}{1981}) \bibinfo{pages}{263--293}.
\bibitem[{Van~Leer(1997)}]{van1997flux}
\bibinfo{author}{B.~Van~Leer},
\newblock \bibinfo{title}{{Flux-Vector Splitting for the Euler Equation}},
\newblock in: \bibinfo{booktitle}{Upwind and High-Resolution Schemes},
  \bibinfo{publisher}{Springer}, \bibinfo{year}{1997}, pp.
  \bibinfo{pages}{80--89}.
\bibitem[{Nickolls and Dally(2010)}]{SIMT}
\bibinfo{author}{J.~Nickolls}, \bibinfo{author}{W.~J. Dally},
\newblock \bibinfo{title}{{The GPU Computing Era}},
\newblock \bibinfo{journal}{IEEE Micro} \bibinfo{volume}{30}
  (\bibinfo{year}{2010}) \bibinfo{pages}{56--69}.
\bibitem[{MPI(2015)}]{MPI}
\bibinfo{title}{{MPI: A Message-Passing Interface Standard}},
  \bibinfo{year}{2015}. \URLprefix
  \url{https://www.mpi-forum.org/docs/mpi-3.1/mpi31-report.pdf},
  \bibinfo{note}{(last accessed on 02/20/20)}.
\bibitem[{ope(2020)}]{openmpi}
\bibinfo{title}{{Open MPI Documentation}}, \bibinfo{year}{2020}. \URLprefix
  \url{https://www.open-mpi.org/doc/}, \bibinfo{note}{(last accessed on
  05/10/20)}.
\bibitem[{mva(2020)}]{mvapich2}
\bibinfo{title}{{MVAPICH: MPI over InfiniBand, Omni-Path, Ethernet/iWARP, and
  RoCE}}, \bibinfo{year}{2020}. \URLprefix
  \url{http://mvapich.cse.ohio-state.edu/userguide/}, \bibinfo{note}{(last
  accessed on 05/10/20)}.
\bibitem[{Hoshino et~al.(2013)Hoshino, Maruyama, Matsuoka, and
  Takaki}]{hoshino2013cuda}
\bibinfo{author}{T.~Hoshino}, \bibinfo{author}{N.~Maruyama},
  \bibinfo{author}{S.~Matsuoka}, \bibinfo{author}{R.~Takaki},
\newblock \bibinfo{title}{Cuda vs openacc: Performance case studies with kernel
  benchmarks and a memory-bound cfd application},
\newblock in: \bibinfo{booktitle}{2013 13th IEEE/ACM International Symposium on
  Cluster, Cloud, and Grid Computing}, \bibinfo{organization}{IEEE},
  \bibinfo{address}{Delft, Netherlands}, \bibinfo{year}{2013}, pp.
  \bibinfo{pages}{136--143}.
\bibitem[{Baig et~al.(2020)Baig, Gao, Teng, Kong, and
  Wang}]{baig2020accelerating}
\bibinfo{author}{F.~Baig}, \bibinfo{author}{C.~Gao}, \bibinfo{author}{D.~Teng},
  \bibinfo{author}{J.~Kong}, \bibinfo{author}{F.~Wang},
\newblock \bibinfo{title}{Accelerating spatial cross-matching on cpu-gpu hybrid
  platform with cuda and openacc.},
\newblock \bibinfo{journal}{Frontiers Big Data} \bibinfo{volume}{3}
  (\bibinfo{year}{2020}) \bibinfo{pages}{14}.
\bibitem[{Artigues et~al.(2020)Artigues, Kormann, Rampp, and
  Reuter}]{artigues2020evaluation}
\bibinfo{author}{V.~Artigues}, \bibinfo{author}{K.~Kormann},
  \bibinfo{author}{M.~Rampp}, \bibinfo{author}{K.~Reuter},
\newblock \bibinfo{title}{Evaluation of performance portability frameworks for
  the implementation of a particle-in-cell code},
\newblock \bibinfo{journal}{Concurrency and Computation: Practice and
  Experience} \bibinfo{volume}{32} (\bibinfo{year}{2020})
  \bibinfo{pages}{e5640}.
\bibitem[{pgi(2019)}]{pgi}
\bibinfo{title}{{PGI Compiler User's Guide}}, \bibinfo{year}{2019}. \URLprefix
  \url{https://www.pgroup.com/resources/docs/19.10/x86/pgi-user-guide/index.htm},
  \bibinfo{note}{(last accessed on 05/10/20)}.
\bibitem[{Hager and Wellein(2010)}]{hager2010introduction}
\bibinfo{author}{G.~Hager}, \bibinfo{author}{G.~Wellein},
  \bibinfo{title}{{Introduction to High Performance Computing for Scientists
  and Engineers}}, \bibinfo{publisher}{CRC Press}, \bibinfo{year}{2010}.
\bibitem[{Farber(2016)}]{farber2016parallel}
\bibinfo{author}{R.~Farber}, \bibinfo{title}{{Parallel Programming with
  OpenACC}}, \bibinfo{publisher}{Newnes}, \bibinfo{year}{2016}.
\bibitem[{Hendrickson and Kolda(2000)}]{hendrickson2000graph}
\bibinfo{author}{B.~Hendrickson}, \bibinfo{author}{T.~G. Kolda},
\newblock \bibinfo{title}{Graph partitioning models for parallel computing},
\newblock \bibinfo{journal}{Parallel computing} \bibinfo{volume}{26}
  (\bibinfo{year}{2000}) \bibinfo{pages}{1519--1534}.
\bibitem[{New(2019)}]{Newriver}
\bibinfo{title}{Newriver}, \bibinfo{year}{2019}. \URLprefix
  \url{https://www.arc.vt.edu/computing/newriver/}, \bibinfo{note}{(last
  accessed on 09/12/20)}.
\bibitem[{Cas(2020)}]{Cascades}
\bibinfo{title}{Cascades}, \bibinfo{year}{2020}. \URLprefix
  \url{https://arc.vt.edu/computing/cascades/}, \bibinfo{note}{(last accessed
  on 09/12/20)}.
\bibitem[{McCall(2017)}]{mccall2017multi}
\bibinfo{author}{A.~J. McCall}, \bibinfo{title}{{Multi-level Parallelism with
  MPI and OpenACC for CFD Applications}}, Master's thesis, Virginia Tech,
  \bibinfo{year}{2017}.
\bibitem[{McCall and Roy(2017)}]{mccall2017multilevel}
\bibinfo{author}{A.~J. McCall}, \bibinfo{author}{C.~J. Roy},
\newblock \bibinfo{title}{{A Multilevel Parallelism Approach with MPI and
  OpenACC for Complex CFD Codes}},
\newblock in: \bibinfo{booktitle}{23rd AIAA Computational Fluid Dynamics
  Conference}, \bibinfo{address}{Denver, CO, USA}, \bibinfo{year}{2017}, p.
  \bibinfo{pages}{3293}.
\bibitem[{Jiayin et~al.(2006)Jiayin, Bo, Yongwei, and
  Guangwen}]{jiayin2006overlapping}
\bibinfo{author}{M.~Jiayin}, \bibinfo{author}{S.~Bo},
  \bibinfo{author}{W.~Yongwei}, \bibinfo{author}{Y.~Guangwen},
\newblock \bibinfo{title}{{Overlapping Communication and Computation in MPI by
  Multithreading}},
\newblock in: \bibinfo{booktitle}{Proc. of International Conference on Parallel
  and Distributed Processing Techniques and Applications},
  \bibinfo{address}{Las Vegas, NEV, USA}, \bibinfo{year}{2006}.
\bibitem[{Vaidyanathan et~al.(2015)Vaidyanathan, Kalamkar, Pamnany, Hammond,
  Balaji, Das, Park, and Jo{\'o}}]{vaidyanathan2015improving}
\bibinfo{author}{K.~Vaidyanathan}, \bibinfo{author}{D.~D. Kalamkar},
  \bibinfo{author}{K.~Pamnany}, \bibinfo{author}{J.~R. Hammond},
  \bibinfo{author}{P.~Balaji}, \bibinfo{author}{D.~Das},
  \bibinfo{author}{J.~Park}, \bibinfo{author}{B.~Jo{\'o}},
\newblock \bibinfo{title}{{Improving Concurrency and Asynchrony in
  Multithreaded MPI Applications using Software Offloading}},
\newblock in: \bibinfo{booktitle}{SC'15: Proceedings of the International
  Conference for High Performance Computing, Networking, Storage and Analysis},
  \bibinfo{organization}{IEEE}, \bibinfo{address}{Austin, TX, USA},
  \bibinfo{year}{2015}, pp. \bibinfo{pages}{1--12}.
\bibitem[{Lu et~al.(2015)Lu, Seo, and Balaji}]{lu2015mpi+}
\bibinfo{author}{H.~Lu}, \bibinfo{author}{S.~Seo}, \bibinfo{author}{P.~Balaji},
\newblock \bibinfo{title}{{MPI+ULT: Overlapping Communication and Computation
  with User-Level Threads}},
\newblock in: \bibinfo{booktitle}{2015 IEEE 17th International Conference on
  High Performance Computing and Communications, 2015 IEEE 7th International
  Symposium on Cyberspace Safety and Security, and 2015 IEEE 12th International
  Conference on Embedded Software and Systems}, \bibinfo{organization}{IEEE},
  \bibinfo{address}{New York, NY, USA}, \bibinfo{year}{2015}, pp.
  \bibinfo{pages}{444--454}.
\bibitem[{Denis and Trahay(2016)}]{denis2016mpi}
\bibinfo{author}{A.~Denis}, \bibinfo{author}{F.~Trahay},
\newblock \bibinfo{title}{{MPI Overlap: Benchmark and Analysis}},
\newblock in: \bibinfo{booktitle}{2016 45th International Conference on
  Parallel Processing (ICPP)}, \bibinfo{organization}{IEEE},
  \bibinfo{address}{Philadelphia, PA, USA}, \bibinfo{year}{2016}, pp.
  \bibinfo{pages}{258--267}.
\bibitem[{Castillo et~al.(2019)Castillo, Jain, Casas, Moreto, Schulz, Beivide,
  Valero, and Bhatele}]{castillo2019optimizing}
\bibinfo{author}{E.~Castillo}, \bibinfo{author}{N.~Jain},
  \bibinfo{author}{M.~Casas}, \bibinfo{author}{M.~Moreto},
  \bibinfo{author}{M.~Schulz}, \bibinfo{author}{R.~Beivide},
  \bibinfo{author}{M.~Valero}, \bibinfo{author}{A.~Bhatele},
\newblock \bibinfo{title}{{Optimizing Computation-Communication Overlap in
  Asynchronous Task-based Programs}},
\newblock in: \bibinfo{booktitle}{Proceedings of the ACM International
  Conference on Supercomputing}, \bibinfo{address}{Washington, DC, USA},
  \bibinfo{year}{2019}, pp. \bibinfo{pages}{380--391}.
\bibitem[{NVIDIA(2019)}]{GPUDirect}
\bibinfo{author}{NVIDIA}, \bibinfo{title}{{NVIDIA GPUDirect}},
  \bibinfo{year}{2019}. \URLprefix
  \url{https://developer.nvidia.com/gpudirect}.
\bibitem[{Mani et~al.(1997)Mani, Ladd, Cain, Bush, Mani, Ladd, Cain, and
  Bush}]{mani1997assessment}
\bibinfo{author}{M.~Mani}, \bibinfo{author}{J.~Ladd},
  \bibinfo{author}{A.~Cain}, \bibinfo{author}{R.~Bush},
  \bibinfo{author}{M.~Mani}, \bibinfo{author}{J.~Ladd},
  \bibinfo{author}{A.~Cain}, \bibinfo{author}{R.~Bush},
\newblock \bibinfo{title}{{An Assessment of One-and Two-Equation Turbulence
  Models for Internal and External Flows}},
\newblock in: \bibinfo{booktitle}{28th Fluid Dynamics Conference},
  \bibinfo{address}{Snowmass Village, CO, USA}, \bibinfo{year}{1997}, p.
  \bibinfo{pages}{2010}.

\end{thebibliography}

\end{document}